\documentclass{article}
\usepackage[T1]{fontenc}
\usepackage{microtype}
\usepackage{graphicx}
\usepackage{booktabs}
\usepackage{amsmath,amssymb}
\usepackage{xcolor}
\usepackage{hyperref}
\usepackage{tikz}
\usetikzlibrary{arrows.meta,positioning,calc,fit,backgrounds}
\usepackage{pgfplots}
\pgfplotsset{compat=1.17}
\usepgfplotslibrary{groupplots}

\usepackage[accepted]{mlsys2025}
\makeatletter
\renewcommand{\mlsys@appearing}{Preprint.}
\makeatother
\usepackage[capitalise]{cleveref}
\ifdefined\pdfinfoomitdate\pdfinfoomitdate=1\fi
\ifdefined\pdftrailerid\pdftrailerid{}\fi
\hypersetup{pdfsubject={Preprint}, pdfauthor={Dezhen Lu, Runguo Li, Zixiang Ni, Yuchen Tang, Xingquan Kong, Jiaxing Yang, Jiarui Guo, Lingxiang Gao}}

\newcommand{\BsixNoneDefaultTailHi}{1.05}
\newcommand{\BsixNoneDefaultTailLo}{0.96}

\newcommand{\BsixNoneDefaultTtftHi}{3.91}
\newcommand{\BsixNoneDefaultTtftLo}{3.38}

\newcommand{\BsixNoneReqTailMs}{17}

\newcommand{\BsixNoneReqTputHi}{466.0}
\newcommand{\BsixNoneReqTputLo}{463.4}
\newcommand{\BsixNoneReqTtft}{2.01}

\newcommand{\BudRowfillBigGbps}{1.26}

\newcommand{\BudRowfillDefaultGbps}{1.40}

\newcommand{\BudTtftCut}{48}
\newcommand{\BurstNoshareTpot}{25.9}
\newcommand{\BurstNoshareTput}{877}
\newcommand{\BurstNoshareTtftFifty}{1.11}
\newcommand{\BurstNoshareTtftNinety}{2.13}
\newcommand{\BurstShareTpot}{22.8}
\newcommand{\BurstShareTput}{1001}
\newcommand{\BurstShareTtftFifty}{0.82}
\newcommand{\BurstShareTtftNinety}{1.63}
\newcommand{\CacheChunkSame}{220/256}
\newcommand{\CacheColdRepeatSame}{128/128}
\newcommand{\CacheHitTokens}{496}
\newcommand{\CacheNoiseFullHi}{90}
\newcommand{\CacheNoiseFullLo}{77}
\newcommand{\CacheNoiseFullMiss}{20}
\newcommand{\CacheNoiseFullSame}{108/128}
\newcommand{\CacheNoiseFullSameEight}{6/8}
\newcommand{\CacheNoiseKindSame}{177/256}
\newcommand{\CacheNoiseN}{128}
\newcommand{\CacheNoisePartHi}{95}
\newcommand{\CacheNoisePartLo}{84}
\newcommand{\CacheNoisePartSame}{116/128}
\newcommand{\CacheSeqFullSame}{6/8}

\newcommand{\CenDtohBytesOne}{8}
\newcommand{\CenHugAnyOpt}{6}
\newcommand{\CenHugAnyPerTokOpt}{2.4}
\newcommand{\CenHugAnyPos}{0}
\newcommand{\CenHugDtohPos}{0.3}
\newcommand{\CenHugDtohSeq}{6.3}

\newcommand{\CenHugItemsOpt}{12}

\newcommand{\CenHugItemsPos}{0}

\newcommand{\CenMoveMainOne}{1004}
\newcommand{\CenMoveOptOne}{29}
\newcommand{\CenPrefillHtodKBNouva}{27}
\newcommand{\CenPrefillHtodNouva}{3.7}

\newcommand{\DiagDouble}{18.4}
\newcommand{\DiagN}{75,776}
\newcommand{\DiagSingle}{0}
\newcommand{\EagerPhaseOne}{0.40}
\newcommand{\EagerPhaseOneTpotMain}{118}
\newcommand{\EagerPhaseOneTpotOpt}{296}

\newcommand{\ExactGainOne}{1.44}
\newcommand{\ExactGainOneTwoEight}{1.14}

\newcommand{\ExactSame}{169/169}

\newcommand{\FaFourHi}{1.03}
\newcommand{\FaFourLo}{0.80}
\newcommand{\FaThreeHi}{0.95}
\newcommand{\FaThreeLo}{0.25}
\newcommand{\FirstRepSlow}{17}
\newcommand{\FiveHitSixteen}{45.7}
\newcommand{\FiveHitSixtyFour}{2.4}
\newcommand{\FiveHitThirtyTwo}{10.7}

\newcommand{\FiveLoopGainOne}{0.99}
\newcommand{\FiveLoopGainThirtyTwo}{1.09}
\newcommand{\FiveLoopSame}{160/169}
\newcommand{\FiveLoopSubsMain}{6.0}
\newcommand{\FiveLoopSubsOpt}{3.0}
\newcommand{\FiveLoopTpotOneMain}{16.2}
\newcommand{\FiveLoopTpotOneOpt}{16.3}

\newcommand{\FivePoolTokens}{24,480}
\newcommand{\FiveRealLoop}{41/41}
\newcommand{\FiveRealSelf}{41/41}

\newcommand{\FullHugPackedDelta}{$-$0.61}

\newcommand{\FullHugPackedLo}{$-$1.5}

\newcommand{\FullHugPackedSe}{0.48}

\newcommand{\FullHugSixteenSame}{1319}

\newcommand{\FullN}{1319}
\newcommand{\FullOuroApproxDelta}{$-$0.15}

\newcommand{\FullOuroApproxNormDelta}{$+$0.76}

\newcommand{\FullOuroApproxNormSe}{0.60}

\newcommand{\FullOuroApproxSe}{0.55}

\newcommand{\FullOuroPackedDelta}{$-$0.99}
\newcommand{\FullOuroPackedDeltaBSixteen}{$-$0.08}

\newcommand{\FullOuroPackedLo}{$-$2.1}

\newcommand{\FullOuroPackedSe}{0.56}

\newcommand{\FullOuroRows}{824}
\newcommand{\FullOuroRowsSixteen}{816}
\newcommand{\GateFullCache}{306}
\newcommand{\GateFullHitKTokens}{293}
\newcommand{\GateFullN}{1319}
\newcommand{\GateFullPos}{307}
\newcommand{\GateFullPosCacheDelta}{$-$0.08}
\newcommand{\GateFullPosCacheSame}{636}

\newcommand{\GateFullSeq}{300}
\newcommand{\GateFullSeqCacheDelta}{$+$0.45}

\newcommand{\GateFullSeqCacheSe}{0.50}
\newcommand{\GateFullSeqPct}{22.7}
\newcommand{\GateFullSeqPosDelta}{$+$0.53}
\newcommand{\GateFullSeqPosSame}{572}
\newcommand{\GateFullSeqPosSe}{0.51}
\newcommand{\GateOuroExact}{62}

\newcommand{\GateOuroSeqDiff}{0}
\newcommand{\GateSmallCache}{21}
\newcommand{\GateSmallN}{87}
\newcommand{\GateSmallPos}{22}

\newcommand{\GateSmallSeq}{23}
\newcommand{\GateSmallSeqCacheDelta}{$-$2.30}

\newcommand{\GateSmallSeqCacheSe}{1.62}
\newcommand{\GateSmallSeqPosDelta}{$-$1.15}
\newcommand{\GateSmallSeqPosDiff}{38}

\newcommand{\GateSmallSeqPosSe}{2.00}
\newcommand{\GhHitSixtyFour}{24.8}
\newcommand{\GhHitThirtyTwo}{79.5}
\newcommand{\GhPoolTokens}{98,336}

\newcommand{\GraphFourNew}{283}
\newcommand{\GraphFourOldHi}{208}
\newcommand{\GraphFourOldLo}{195}

\newcommand{\GraphSixteenRuns}{23}
\newcommand{\GraphSixteenSlow}{3}

\newcommand{\GraphSlowFallbacks}{252}

\newcommand{\GsmAllCand}{62}
\newcommand{\GsmAllDelta}{$-$2.30}

\newcommand{\GsmAllSe}{1.62}
\newcommand{\GsmAllSeqDiff}{23}
\newcommand{\GsmN}{87}
\newcommand{\GsmNoneCand}{64}

\newcommand{\GsmNoneSeqDiff}{0}
\newcommand{\GsmNormCand}{61}
\newcommand{\GsmNormDelta}{$-$3.45}
\newcommand{\GsmNormRef}{64}
\newcommand{\GsmNormSe}{2.56}
\newcommand{\GsmNormSeqDiff}{29}
\newcommand{\HsplitFullEight}{1.92}
\newcommand{\HsplitFullOne}{1.59}
\newcommand{\HsplitFullThirtyTwo}{2.57}
\newcommand{\HsplitKernEight}{1.07}
\newcommand{\HsplitKernOne}{1.00}
\newcommand{\HsplitKernThirtyTwo}{1.03}

\newcommand{\HsplitRestOne}{1.59}
\newcommand{\HsplitRestThirtyTwo}{2.49}
\newcommand{\HugBaseDriftOne}{17}

\newcommand{\HugFiveKvTokens}{1328}
\newcommand{\HugFiveTputCEight}{17.8}
\newcommand{\HugFiveTputCSixtyFour}{16.7}
\newcommand{\HugFullMBPerToken}{5.41}

\newcommand{\HugFuseHi}{1.11}
\newcommand{\HugFuseLo}{1.07}

\newcommand{\HugGateN}{70}
\newcommand{\HugGatePacked}{21}
\newcommand{\HugGateRows}{20}

\newcommand{\HugLiveMBPerToken}{2.79}
\newcommand{\HugMainPlateau}{65}

\newcommand{\HugOptMax}{1.83}
\newcommand{\HugOptMin}{1.49}

\newcommand{\HugPackedPfiveGainThirtyTwo}{1.33}
\newcommand{\HugPackedPfiveSame}{36/41}
\newcommand{\HugPackedPtwokGainThirtyTwo}{2.08}
\newcommand{\HugPackedPtwokTtftPacked}{18.1}
\newcommand{\HugPackedPtwokTtftRows}{41.9}

\newcommand{\HugSessBaseSixtyFour}{14.0}

\newcommand{\HugSessGain}{2.72}
\newcommand{\HugSessGainB}{3.12}
\newcommand{\HugSessGainExact}{2.34}
\newcommand{\HugSessGainNopc}{1.16}
\newcommand{\HugSessIsoGain}{2.12}

\newcommand{\HugSessPcHit}{78}

\newcommand{\HugSessPdThirteenGain}{2.89}
\newcommand{\HugSessReqs}{59}

\newcommand{\HugSessTtftRatio}{15}
\newcommand{\HugSubsMain}{17.7}
\newcommand{\HugSubsOpt}{3.0}
\newcommand{\IsoExactSame}{208/208}
\newcommand{\IsoMainOptSame}{202/208}
\newcommand{\IsoMainSelfSame}{208/208}

\newcommand{\KernExactCut}{49}
\newcommand{\KernExactGpuMs}{8.91}
\newcommand{\KernExactPerToken}{2885}
\newcommand{\KernOffGpuMs}{13.05}
\newcommand{\KernOffPerToken}{5689}

\newcommand{\KfApproxMax}{1.67}
\newcommand{\KfApproxMin}{1.23}

\newcommand{\KfApproxSame}{894/1014}

\newcommand{\KfExactEight}{1.59}
\newcommand{\KfExactMax}{1.59}
\newcommand{\KfExactMin}{1.11}
\newcommand{\KfExactOne}{1.55}
\newcommand{\KfExactOneTwoEight}{1.11}
\newcommand{\KfExactSame}{1014/1014}
\newcommand{\KfExactThirtyTwo}{1.51}

\newcommand{\KfThreeMax}{1.30}
\newcommand{\KfThreeMin}{0.97}

\newcommand{\KfVfourExactMax}{1.50}
\newcommand{\KfVfourExactMin}{1.03}
\newcommand{\KfVfourExactSame}{507/507}
\newcommand{\MathHugPackedDelta}{$-$0.20}
\newcommand{\MathHugPackedLo}{$-$1.4}

\newcommand{\MathHugPackedSe}{0.61}
\newcommand{\MathN}{496}
\newcommand{\MathOuroPackedDelta}{$+$0.40}
\newcommand{\MathOuroPackedLo}{$-$2.0}

\newcommand{\MathOuroPackedSe}{1.21}
\newcommand{\MathOuroUnfusedSame}{496}
\newcommand{\NcuDramMax}{0.2}

\newcommand{\NcuMaxUs}{3.8}
\newcommand{\NcuMinUs}{2.4}

\newcommand{\NoisePosHiEight}{1.04}
\newcommand{\NoisePosHiOne}{1.00}
\newcommand{\NoisePosHiThirtyTwo}{1.09}
\newcommand{\NoisePosLoEight}{1.01}
\newcommand{\NoisePosLoOne}{0.92}
\newcommand{\NoisePosLoThirtyTwo}{1.04}
\newcommand{\OnAbortRuns}{2}
\newcommand{\OnAlpacaOutMean}{58}
\newcommand{\OnHugNlGainMax}{4.2}
\newcommand{\OnHugNlGainMin}{2.9}
\newcommand{\OnHugOffAlpacaGain}{1.60}
\newcommand{\OnHugOffShareGPTGain}{2.01}

\newcommand{\OnHugRateMax}{1.5}

\newcommand{\OnHugRateN}{5}

\newcommand{\OnHugTtftWorseN}{2}
\newcommand{\OnOuroBaseKnee}{4.3}
\newcommand{\OnOuroBaseTwo}{0.054}
\newcommand{\OnOuroNlGainMax}{5.6}

\newcommand{\OnOuroNlSix}{2.3}
\newcommand{\OnOuroOffAlpacaGain}{1.29}
\newcommand{\OnOuroOffShareGPTGain}{1.71}
\newcommand{\OnOuroOptFour}{0.041}

\newcommand{\OnOuroPlateauBaseHi}{386}
\newcommand{\OnOuroPlateauBaseLo}{343}
\newcommand{\OnOuroPlateauOptHi}{390}
\newcommand{\OnOuroPlateauOptLo}{308}

\newcommand{\OnOuroRateN}{6}
\newcommand{\OnOuroSatRate}{8}

\newcommand{\OnOuroTtftWorseN}{1}
\newcommand{\OnPlainRuns}{10}
\newcommand{\OnShareKept}{911}
\newcommand{\OnShareRows}{2000}
\newcommand{\OneBenchHi}{4.4}

\newcommand{\OneBenchLo}{2.0}
\newcommand{\OneCases}{156}

\newcommand{\OneHugHi}{1.08}
\newcommand{\OneHugLo}{1.06}

\newcommand{\OneOuroLongHi}{1.17}

\newcommand{\OneOuroShortHi}{1.18}
\newcommand{\OneOuroShortLo}{1.13}

\newcommand{\OuroBigLongMax}{3.3}
\newcommand{\OuroBigLongMin}{2.2}
\newcommand{\OuroBigLongSame}{67/120}
\newcommand{\OuroBigOffGain}{1.68}
\newcommand{\OuroBigShortMax}{2.3}
\newcommand{\OuroBigShortMin}{2.0}
\newcommand{\OuroBigShortSame}{272/315}

\newcommand{\PackCapacityRatio}{1.88}
\newcommand{\PackGainSixtyFour}{1.16}
\newcommand{\PackGainThirtyTwo}{1.14}
\newcommand{\PackPackedTokens}{22736}

\newcommand{\PackSame}{233/233}
\newcommand{\PackUnpackedTokens}{12080}
\newcommand{\PackedPtwokEight}{2.42}
\newcommand{\PackedPtwokSame}{37/40}
\newcommand{\PackedPtwokThirtyTwo}{2.00}
\newcommand{\PackedPtwokTtftCut}{61}

\newcommand{\PdHugRatio}{0.516}

\newcommand{\PdMainCompute}{2.87}
\newcommand{\PdMainGbps}{1.44}
\newcommand{\PdMainTail}{2.85}

\newcommand{\PdMainTputThirtyTwo}{129.0}

\newcommand{\PdMainXferMs}{44.5}
\newcommand{\PdOptCompute}{0.64}

\newcommand{\PdOptGainThirtyTwo}{2.28}

\newcommand{\PdOptTputThirtyTwo}{294.4}

\newcommand{\PdOptXferMs}{14.5}
\newcommand{\PdOuroGBPerRequest}{1.6}
\newcommand{\PdOuroKBPerToken}{786}
\newcommand{\PdRowsCompute}{2.69}

\newcommand{\PdRowsTputThirtyTwo}{157.5}

\newcommand{\PdfBudgetGain}{1.7}
\newcommand{\PdfGain}{3.3}
\newcommand{\PdfGainPeer}{3.1}
\newcommand{\PdfLongEightG}{86}
\newcommand{\PdfLongGBPerRequest}{6.4}
\newcommand{\PdfLongMain}{13.8}
\newcommand{\PdfLongMainTtft}{33.7}
\newcommand{\PdfLongNoneHi}{89}
\newcommand{\PdfLongNoneLo}{79}
\newcommand{\PdfLongNoneTtftHi}{4.5}
\newcommand{\PdfLongNoneTtftLo}{3.8}
\newcommand{\PdfLongTailEightMs}{43}
\newcommand{\PdfLongTailNoneMs}{41}
\newcommand{\PdfLongTailOldHi}{0.27}
\newcommand{\PdfLongTailOldLo}{0.08}
\newcommand{\PdfMainHi}{147}
\newcommand{\PdfMainLo}{143}

\newcommand{\PdfMainTtft}{11.0}

\newcommand{\PdfNoneLo}{479}

\newcommand{\PdfNoneTtft}{2.0}
\newcommand{\PdfOldHi}{308}
\newcommand{\PdfOldLo}{271}

\newcommand{\PdfPeerFastHi}{457}
\newcommand{\PdfPeerFastLo}{450}
\newcommand{\PdfPeerFastTpotMs}{21}

\newcommand{\PdfPeerSlow}{205}
\newcommand{\PdfPeerSlowTpotMs}{116}

\newcommand{\PdfTailMain}{2.8}
\newcommand{\PdfTailNoneMs}{16}
\newcommand{\PdfTailOld}{1.0}
\newcommand{\PdfTestsMain}{52}
\newcommand{\PdfTestsPr}{56}
\newcommand{\PdfTwoMain}{236}

\newcommand{\PdfTwoNoneHi}{617}
\newcommand{\PdfTwoNoneLo}{610}
\newcommand{\PdfTwoNoneSixtyFourHi}{751}
\newcommand{\PdfTwoNoneSixtyFourLo}{460}
\newcommand{\PdfTwoOldHi}{450}
\newcommand{\PdfTwoOldLo}{446}

\newcommand{\PfGainEight}{1.01}
\newcommand{\PfGainSixtyFour}{1.28}
\newcommand{\PfGainThirtyTwo}{1.21}
\newcommand{\PfSame}{309/312}
\newcommand{\PfTtftPackSixtyFour}{5.5}
\newcommand{\PfTtftUnpSixtyFour}{10.8}
\newcommand{\PrefillSlowMainThirtyTwo}{2.32}
\newcommand{\PrefillSlowOptThirtyTwo}{3.08}
\newcommand{\ProbeCases}{132}
\newcommand{\ProjAllHi}{1.12}
\newcommand{\ProjAllLo}{1.06}
\newcommand{\ProjDecodeCut}{12}

\newcommand{\ProjLongHi}{1.08}
\newcommand{\ProjLongLo}{1.07}

\newcommand{\ProjShortHi}{1.12}
\newcommand{\ProjShortLo}{1.06}
\newcommand{\ProjSweepTop}{8192}
\newcommand{\RelAllMax}{3.2}
\newcommand{\RelAllMin}{1.6}
\newcommand{\RelBHugMax}{2.7}
\newcommand{\RelBHugMin}{1.5}
\newcommand{\RelBOuroLongMax}{3.8}
\newcommand{\RelBOuroLongMin}{3.5}
\newcommand{\RelBOuroShortMax}{2.3}
\newcommand{\RelBOuroShortMin}{2.0}
\newcommand{\RelFallbacks}{0}

\newcommand{\RelHugMax}{3.16}
\newcommand{\RelHugMin}{1.77}
\newcommand{\RelHugOne}{1.77}

\newcommand{\RelHugSame}{271/315}
\newcommand{\RelHugTtftMainSixtyFour}{22.7}
\newcommand{\RelHugTtftOptSixtyFour}{4.0}

\newcommand{\RelN}{2}
\newcommand{\RelOldHugMax}{2.21}
\newcommand{\RelOldHugMin}{1.43}
\newcommand{\RelOldOuroLongMax}{2.54}
\newcommand{\RelOldOuroLongMin}{2.33}
\newcommand{\RelOldOuroShortMax}{1.57}
\newcommand{\RelOldOuroShortMin}{1.45}
\newcommand{\RelOuroLongMax}{2.04}
\newcommand{\RelOuroLongMin}{1.91}
\newcommand{\RelOuroLongSame}{109/120}
\newcommand{\RelOuroMax}{2.15}
\newcommand{\RelOuroMin}{1.65}
\newcommand{\RelOuroShortMax}{2.15}
\newcommand{\RelOuroShortMin}{1.65}
\newcommand{\RelOuroShortSame}{285/315}
\newcommand{\RelPrevHugMax}{2.62}
\newcommand{\RelPrevHugMin}{1.56}
\newcommand{\RelPrevOuroLongMax}{2.92}
\newcommand{\RelPrevOuroLongMin}{2.79}
\newcommand{\RelPrevOuroShortMax}{1.84}
\newcommand{\RelPrevOuroShortMin}{1.59}
\newcommand{\RelSixOuroShortMax}{2.06}
\newcommand{\RelSixOuroShortMin}{1.75}
\newcommand{\RowsHugHi}{1.05}
\newcommand{\RowsHugPrefillCut}{14}
\newcommand{\RowsOuroLongHi}{1.06}

\newcommand{\RowsOuroPrefillCut}{29}

\newcommand{\ScopeRepeatThirtyTwo}{0.84}
\newcommand{\SessFirstSlowMax}{33}

\newcommand{\SessMainTtftSixtyFour}{1.41}

\newcommand{\SessOptTtftSixtyFour}{6.41}

\newcommand{\ShareN}{2}

\newcommand{\SiluMismatch}{0}

\newcommand{\SpfOptEight}{1.31}
\newcommand{\SpfOptOne}{1.52}

\newcommand{\SpkEightOne}{1.55}

\newcommand{\SpkFixedThirtyTwo}{0.30}
\newcommand{\SpkStepsPlain}{408}
\newcommand{\SpkStepsZero}{151}
\newcommand{\SpkZeroHi}{1.11}
\newcommand{\SpkZeroLo}{1.06}

\newcommand{\SqEightFour}{1.26}
\newcommand{\SqEightLowSame}{28/28}
\newcommand{\SqEightOne}{1.52}
\newcommand{\SqEightSame}{125/128}

\newcommand{\SqEightThirtyTwo}{1.01}
\newcommand{\SqEightTwo}{1.37}
\newcommand{\SqFallbacks}{0}
\newcommand{\SqFixedEight}{1.13}
\newcommand{\SqFixedFour}{1.13}
\newcommand{\SqFixedGraphs}{85}
\newcommand{\SqFixedOne}{1.52}
\newcommand{\SqFixedSixteen}{1.05}
\newcommand{\SqFixedThirtyTwo}{0.75}
\newcommand{\SqFixedTwo}{1.37}

\newcommand{\SqZeroHighHi}{1.04}
\newcommand{\SqZeroHighLo}{1.01}
\newcommand{\SqZeroLowHi}{1.08}
\newcommand{\SqZeroLowLo}{1.07}

\newcommand{\SqZeroSame}{126/128}

\newcommand{\SsBaseSixteenTtftSixtyFour}{1.48}

\newcommand{\SsBaseTpotSixtyFour}{0.99}

\newcommand{\SsBaseTtftSixtyFour}{1.95}

\newcommand{\SsFinalGainSixtyFour}{3.21}
\newcommand{\SsFinalGainThirtyTwo}{1.91}
\newcommand{\SsFinalTtftBaseSixtyFour}{1.98}

\newcommand{\SsFinalTtftRelSixtyFour}{0.15}

\newcommand{\SsGraphGainSixtyFour}{1.10}
\newcommand{\SsGraphGainThirtyTwo}{1.15}
\newcommand{\SsPackedGainSixtyFour}{2.79}
\newcommand{\SsPackedTpotSixtyFour}{0.14}

\newcommand{\SsPackedTtftNinetySixtyFour}{4.25}

\newcommand{\SsPackedTtftSixtyFour}{0.94}

\newcommand{\SsRelGainSixtyFour}{2.45}
\newcommand{\SsRelGainThirtyTwo}{1.58}
\newcommand{\SsRelTpotSixtyFour}{0.32}

\newcommand{\SsRelTtftNinetySixtyFour}{1.15}

\newcommand{\SsRelTtftSixtyFour}{0.25}

\newcommand{\SsRowsTpotSixtyFour}{0.43}

\newcommand{\SsRowsTtftSixtyFour}{3.44}

\newcommand{\SsShareGainSixtyFour}{0.88}
\newcommand{\SsShareGainThirtyTwo}{0.96}
\newcommand{\SsSixGainSixtyFour}{3.00}
\newcommand{\SsSixGainThirtyTwo}{1.87}

\newcommand{\SubsOuroMain}{5.9}
\newcommand{\SubsOuroOpt}{3.0}

\newcommand{\UcxNotcpXferMs}{16.9}

\newcommand{\UcxPeerXferMs}{20.0}

\newcommand{\UcxRowsXferMs}{51.7}
\newcommand{\UvaN}{2}
\newcommand{\UvaNouvaEight}{0.88}
\newcommand{\UvaNouvaThirtyTwo}{0.93}
\newcommand{\UvaPackedEight}{0.98}
\newcommand{\UvaPackedThirtyTwo}{1.19}
\newcommand{\UvaRepeatEight}{0.98}
\newcommand{\UvaRepeatThirtyTwo}{1.10}
\newcommand{\VtwoHugMax}{2.60}
\newcommand{\VtwoHugMin}{1.85}

\newcommand{\VtwoOuroShortMax}{1.45}
\newcommand{\VtwoOuroShortMin}{1.33}

\newcommand{\X}{$\times$}
\newcommand{\rung}[1]{\textsc{r#1}}
\newcommand{\appref}[1]{\cref{#1}}
\newcommand{\Appref}[1]{\Cref{#1}}

\mlsystitlerunning{What Sets the Serving Speed of Looped Language Models?}

\begin{document}

\twocolumn[
\mlsystitle{What Sets the Serving Speed of Looped Language Models?\\ Measurements and Optimizations Graded by Exactness}

\begin{mlsysauthorlist}
\mlsysauthor{Dezhen Lu}{tue}
\mlsysauthor{Runguo Li}{uiuc}
\mlsysauthor{Zixiang Ni}{xjtu}
\mlsysauthor{Yuchen Tang}{zstu}
\mlsysauthor{Xingquan Kong}{swpu}
\mlsysauthor{Jiaxing Yang}{fudan}
\mlsysauthor{Jiarui Guo}{nankai}
\mlsysauthor{Lingxiang Gao}{csuft}
\end{mlsysauthorlist}
\mlsysaffiliation{tue}{University of T\"ubingen, T\"ubingen, Germany}
\mlsysaffiliation{uiuc}{University of Illinois Urbana-Champaign, Urbana, IL, USA}
\mlsysaffiliation{xjtu}{Xi'an Jiaotong University, Xi'an, China}
\mlsysaffiliation{zstu}{Zhejiang Sci-Tech University, China}
\mlsysaffiliation{swpu}{Southwest Petroleum University, China}
\mlsysaffiliation{fudan}{Fudan University, Shanghai, China}
\mlsysaffiliation{nankai}{Nankai University, Tianjin, China}
\mlsysaffiliation{csuft}{Swan College, Central South University of Forestry and Technology, Changsha, China}
\mlsyscorrespondingauthor{Dezhen Lu}{dezhen.lu@student.uni-tuebingen.de}

\mlsyskeywords{LLM serving, looped language models, KV cache, kernel fusion, determinism}

\vskip 0.3in

\begin{abstract}
Looped language models apply one block of layers many times per token, so a serving engine pays every per-layer
overhead once per loop; we measure serving for Ouro and Huginn-0125 in an open-source looped-model engine and grade
our optimizations from bitwise-identical changes to those that require an accuracy check. At batch 1, fusing all
loop submissions leaves Ouro throughput at \FiveLoopGainOne\X; the \KernOffPerToken{} small kernels per token are
the real cost, and fusing them behind \emph{rounding fences} raises throughput
\KfExactMin--\KfExactMax\X{} while \KfExactSame{} requests remain identical. An approximate fusion is no faster
and misses a one-point GSM8K rule on 87 questions (61 and 62 correct versus 64), so we do not ship it. Huginn's
released defaults avoid dead KV slots, reuse prefix-cache pages, and serve \RelHugMin--\RelHugMax\X{} faster than
the base engine; Ouro reaches \RelOuroMin--\RelOuroMax\X{} and agent sessions
\SsFinalGainThirtyTwo--\SsFinalGainSixtyFour\X{}, while Huginn's noise and prefix-reuse changes stay within one point on all
\GateFullN{} Huginn test questions; Huginn's packed prefill, \FullHugPackedDelta{} points on the same questions, meets that rule only on its point estimate. Position-keyed noise makes CPU cache reuse exact but still leaves \CacheNoiseFullMiss{} of \CacheNoiseN{} GPU
full hits different, while loop fusion delays prefill admission and a 64-graph cache prevents eager fallbacks.
Under prefill--decode disaggregation ExactLoop reaches up to \PdfGain\X{} the base throughput with peer error handling
disabled, and \PdfGainPeer\X{} in two of three runs with it enabled. Its transfer budget, sized to hold one request's
KV in flight, cuts the commit tail from about one second under the earlier budget to \PdfTailNoneMs~ms (base:
\PdfTailMain~s), removing the polling rounds that the engine's interval would otherwise impose.

\end{abstract}
]

\printAffiliationsAndNotice{}

\section{Introduction}
\label{sec:intro}

Looped, or recurrent-depth, language models reuse one block of transformer layers several times for each
token. Huginn-0125 runs a four-layer block 32 times between a short prelude and coda~\cite{geiping2025huginn};
Ouro runs all of its layers four times and can exit early~\cite{zhu2025ouro}. Depth becomes a run-time knob,
which makes these models attractive for reasoning at a fixed parameter count. For a serving engine a looped
model looks like a much deeper model whose layers share weights. The natural first optimization is to fuse the
loops of a decode step into one GPU submission, which removes host round trips that grow with depth.

We tried this on vllm-rlt~\cite{vllmrlt}, an open-source engine built for looped models (main branch at commit
66920da), and it did almost
nothing at low load. On an RTX~5090, fusing Ouro's four loops halved the submissions per token
(\FiveLoopSubsMain{} to \FiveLoopSubsOpt) and left batch-1 throughput at \FiveLoopGainOne\X; it helped only at
higher load (\FiveLoopGainThirtyTwo\X{} at 32 requests). The loops are not the problem. The problem is that every
per-layer cost is paid once per loop, so small costs that a conventional engine can ignore become large. We
found two that matter most.

\paragraph{Dead KV slots.} The engine keeps a separate KV plane for each recurrence depth, so that a token at
depth $d$ attends to keys computed at depth $d$. Each plane reserves a slot for every layer. In Huginn only the
four recurrent layers write deeper planes; the four prelude and coda layers write plane 0 only. Of the
$8\times32=256$ layer slots a token reserves, 124 are never written. On a GH200 the base engine therefore stops
scaling near \HugMainPlateau{} tokens/s, because only a few dozen requests fit in the KV pool. The same layout
breaks prefix caching: the cache publishes a prefix only when every layer of every plane has written it, which
never happens for Huginn.

\paragraph{Kernel count.} At batch 1, an Ouro token costs \KernOffPerToken{} GPU kernels on a GH200, most of them
small elementwise and reduction kernels from RMSNorm, rotary embedding (RoPE) and the gated activation, each
repeated for 24 layers and four loops. CUDA graphs remove their launch cost but not their minimum run time.
Fusing them is the obvious remedy, but fused kernels round differently, and in a recurrent model a one-ulp
difference passes through 96 layer applications. Our first fused kernels changed GSM8K answers.

\paragraph{The exactness ladder.} An optimization for a deployed model is easier to adopt the less it changes
outputs. \Cref{tab:ladder} records what each ExactLoop change preserves at two levels: the arithmetic applied to
the same inputs and batch, and the tokens observed after deployment. A \rung{0} change is bitwise identical under
the same batch composition. It may change capacity or scheduling, which can in turn expose the batch-size-dependent
arithmetic called \rung{1}; an end-to-end comparison then contains both effects. \rung{1} changes preserve each
row's operations but change the rows that share a batch, so GEMM algorithms can flip bits
~\cite{he2025nondeterminism}. \rung{2} changes target the same sampling distribution with different random
numbers. \rung{3} changes alter arithmetic and need a task-level accuracy check. Rungs 0 and 1 produce the Huginn
high-concurrency and Ouro batch-1 gains. Long-prompt and session gains come mainly from rungs 2 and 3, while the
PD gain is split between rung-3 packed prefill and a rung-0 transfer budget.

\paragraph{Contributions.}
\begin{itemize}\setlength{\itemsep}{1pt}\setlength{\parskip}{0pt}\setlength{\topsep}{2pt}
\item A characterization of a looped-model engine that traces Huginn's throughput ceiling to dead KV slots,
Ouro's batch-1 latency to kernel count, and Huginn's zero prefix-cache hit rate to the publication rule and to
order-dependent recurrent-state noise (\cref{sec:where}).
\item Plane packing, which stores only live layer slots and holds \PackCapacityRatio\X{} more Huginn tokens in the
same pool with bitwise-identical outputs under the same batch composition (\cref{sec:packing}).
\item Exact kernel fusion: fused RMSNorm, RoPE and SiLU kernels that are bitwise identical to PyTorch because the
row sum keeps PyTorch's order and \emph{rounding fences}, inline PTX conversions the compiler cannot see through,
pin every rounding point on the tested PyTorch, CUDA and GPU configurations (\cref{sec:fusion}).
\item Position-keyed recurrent-state noise, which makes Huginn's random initial state a function of
(seed, position, channel), so a cached prefix sees the same noise as a cold run independent of draw order
(\cref{sec:noise}).
\item An evaluation that reports token-level matches alongside throughput and latency, covers two GSM8K check
sizes, real prompts under Poisson arrivals and a third model, and documents four negative results: approximate fusion, speculation at high load, eager-mode kernel
launch cost, and a KV transfer paced by engine polling (\cref{sec:eval,sec:lessons}).
\end{itemize}
The supplementary appendix lists every GPU batch we ran, the question it asked and its outcome, including the
batches that produced no data and the ones that refuted our own explanations.

Most of the individual techniques have precedents: paged KV layouts, kernel fusion, counter-based random numbers and
transfer budgets. What this paper adds is their measured effect on looped models, and an exactness grading that
decides which of them can be shipped without an accuracy check.

\begin{table}[t]
\caption{The exactness ladder. Each ExactLoop change, the property it preserves, and the evidence. ``Identical''
counts requests whose generated token sequences match the same configuration without the change.}
\label{tab:ladder}
\centering\small
\setlength{\tabcolsep}{3pt}
\begin{tabular}{@{}llp{3.6cm}@{}}
\toprule
rung & change & evidence \\
\midrule
\rung{0} & plane packing & Huginn \PackSame{} identical with unchanged batches; \PfSame{} when more requests fit (the latter also changes \rung{1} batch composition) \\
\rung{0} & exact kernel set & Ouro \ExactSame{}, \IsoExactSame{} and \KfExactSame{} identical; GSM8K \GsmN{} questions, \GateOuroSeqDiff{} sequences differ \\
\rung{0} & PD fixes, transfer budget & transfer layout and timing only; PD tests pass \\
\rung{0} & graph budget, eviction & which steps replay a graph; graph tests pass \\
\rung{1} & adaptive drafts & \SqEightSame{} identical at 32 requests, \SqEightLowSame{} at $C{\le}4$ \\
\rung{1} & loop fusion & batch 1: GSM8K \GsmNoneSeqDiff{} of \GsmN{} differ; at $C{\ge}8$, Ouro \IsoMainOptSame{} identical \\
\rung{2} & position-keyed noise & CPU: cached = cold; GPU: full hits \CacheNoiseFullSame, partial hits \CacheNoisePartSame; GSM8K \GateFullSeqPosDelta{} points (s.e.\ \GateFullSeqPosSe, \GateFullN{} questions) \\
\rung{3} & packed prefill & Ouro \PackedPtwokSame{}, Huginn \HugPackedPfiveSame{} identical; GSM8K: Ouro \FullOuroPackedDelta, Huginn \FullHugPackedDelta{} points \\
\rung{3} & approximate fusion & not shipped: not bitwise, no faster (\cref{sec:gates}) \\
\bottomrule
\end{tabular}
\end{table}

\section{Background}
\label{sec:background}

\paragraph{Looped models.} Ouro-1.4B has 24 transformer layers that run four times per token; a learned gate
can stop a token early~\cite{zhu2025ouro}. We run it at a fixed depth of four. Huginn-0125 embeds a token, runs
two prelude layers, initializes a latent state with truncated normal noise, applies a four-layer recurrent block
32 times to the state and the prelude output, and decodes the result with two coda
layers~\cite{geiping2025huginn}. Earlier weight-shared models include Universal
Transformers~\cite{dehghani2019universal}; early-exit models such as CALM~\cite{schuster2022calm} and
LayerSkip~\cite{elhoushi2024layerskip} also vary depth per token.

\paragraph{The engine.} vllm-rlt extends paged KV caching~\cite{kwon2023vllm} and continuous
batching~\cite{yu2022orca} to looped models and implements runtime ideas from continuous depth
batching~\cite{schwethelm2026cdb}. A request moves through PREFILL, PRELUDE, RECURRENT and CODA phases,
and the scheduler forms each batch from rows in one phase. In the LAST\_EXITED layout every recurrence depth
has its own KV plane, a token at depth $d$ attends to plane $d$, and a row that exits early copies its last
plane forward. The engine supports asynchronous scheduling, CUDA graphs for decode, Triton~\cite{tillet2019triton}
paged attention, a prefix cache, self-speculative decoding with a shallower draft~\cite{zhang2024draftverify},
and prefill--decode (PD) disaggregation~\cite{zhong2024distserve,patel2024splitwise} that writes KV pages into
the decode engine's pool with NIXL~\cite{nixl}. Prefill runs one row at a time by default; a packed,
variable-length prefill path exists but is opt-in.

\section{Per-Loop Overheads, Not Loop Arithmetic, Set the Speed}
\label{sec:where}

We profiled the base engine with random-token burst workloads (all requests arrive at once, greedy decoding,
fixed depth) and with multi-turn agent-session traces (\cref{sec:setup}).

\paragraph{Loop submissions do not limit batch-1 decode.} On the RTX~5090 an Ouro token at batch 1 took
\FiveLoopTpotOneMain{}~ms. Fusing its loops cut the submissions per token from \FiveLoopSubsMain{} to
\FiveLoopSubsOpt{} and the per-step host time several-fold, yet the time per output token (TPOT) stayed at
\FiveLoopTpotOneOpt{}~ms. A kernel profile on the GH200 explains why: \KernOffPerToken{} kernels and
\KernOffGpuMs{}~ms of GPU time per token. With graphs the host no longer launches these kernels, but each still
runs for a few microseconds. Each RMSNorm is eight kernels in PyTorch~\cite{paszke2019pytorch}, four norms per
layer.

\paragraph{Dead KV slots cap Huginn's concurrency.} For each Huginn token, upstream reserves \HugFullMBPerToken{}~MB of KV, of
which \HugLiveMBPerToken{}~MB is ever written. On the RTX~5090 the pool held \HugFiveKvTokens{} tokens, two
requests of our burst, and throughput stayed at \HugFiveTputCEight{} tokens/s from 8 to 64 concurrent requests
(\HugFiveTputCSixtyFour{} at 64). On a GH200 the pool is larger but the effect is the same: the base engine's
throughput stayed near \HugMainPlateau{} tokens/s from 32 requests on (\cref{tab:release}).

\paragraph{Cache hits follow KV capacity, and Huginn publishes no prefix.} Ouro sessions on the GH200 reused
\GhHitThirtyTwo\% of prompt tokens from the prefix cache with 32 sessions but \GhHitSixtyFour\% with 64, whose
working set exceeds the \GhPoolTokens-token pool. On the RTX~5090, whose pool holds \FivePoolTokens{} tokens,
the hit rate fell from \FiveHitSixteen\% to \FiveHitThirtyTwo\% to \FiveHitSixtyFour\% for 16, 32 and 64
sessions. Huginn never hit at all: the cache publishes a page only when every layer of every plane has written
it, and Huginn's boundary layers never write planes 1--31.

\paragraph{Prefill runs one row at a time.} With 2048-token Ouro prompts, packed prefill was \PackedPtwokEight\X{}
and \PackedPtwokThirtyTwo\X{} faster than the per-row default at 8 and 32 requests, with \PackedPtwokTtftCut\%
lower time to first token (TTFT) at 32.

\paragraph{KV transfer stalls behind prefill.} In a one-prefill, one-decode deployment on two GPUs of one GH200
node, the base engine at 32 requests spent a median of \PdMainTail~s per request between the end of prefill compute and the
last KV chunk leaving (the \emph{commit tail}), as long as the prefill compute itself (\PdMainCompute~s). The
transfer ran at \PdMainGbps~GB/s.

\section{ExactLoop: Removing the Overheads, Graded by Exactness}
\label{sec:design}

\subsection{Packing live slots changes addresses, not values (\rung{0})}
\label{sec:packing}

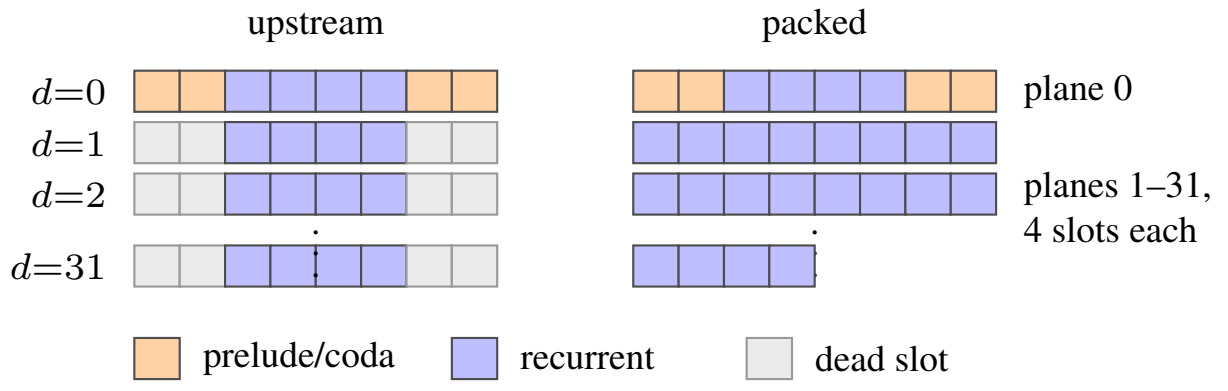
\begin{figure}[t]
\centering
\begin{tikzpicture}[x=0.3cm,y=0.34cm,font=\scriptsize]
  \tikzset{live/.style={draw=black!70,fill=blue!25},dead/.style={draw=black!40,fill=black!8},
           pre/.style={draw=black!70,fill=orange!35}}
  \node[anchor=south] at (4,5.0) {upstream};
  \foreach \p/\y in {0/4,1/3,2/2,31/0.6} {
    \node[anchor=east] at (-0.2,\y+0.4) {$d{=}\p$};
    \foreach \s in {0,...,7} {
      \pgfmathsetmacro{\isb}{(\s<2 || \s>5) ? 1 : 0}
      \ifnum\p=0
        \ifdim\isb pt=1pt \draw[pre] (\s,\y) rectangle ++(1,0.8); \else \draw[live] (\s,\y) rectangle ++(1,0.8); \fi
      \else
        \ifdim\isb pt=1pt \draw[dead] (\s,\y) rectangle ++(1,0.8); \else \draw[live] (\s,\y) rectangle ++(1,0.8); \fi
      \fi
    }
  }
  \node at (4,1.55) {$\vdots$};
  \node[anchor=south] at (15,5.0) {packed};
  \foreach \s in {0,...,7} {
    \pgfmathsetmacro{\isb}{(\s<2 || \s>5) ? 1 : 0}
    \ifdim\isb pt=1pt \draw[pre] (11+\s,4) rectangle ++(1,0.8); \else \draw[live] (11+\s,4) rectangle ++(1,0.8); \fi
  }
  \foreach \b/\y in {1/3,2/2} { \foreach \s in {0,...,7} { \draw[live] (11+\s,\y) rectangle ++(1,0.8); } }
  \node at (15,1.55) {$\vdots$};
  \foreach \s in {0,...,3} { \draw[live] (11+\s,0.6) rectangle ++(1,0.8); }
  \node[anchor=west] at (19.2,4.4) {plane 0};
  \node[anchor=west] at (19.2,2.5) {planes 1--31,};
  \node[anchor=west] at (19.2,1.7) {4 slots each};
  \draw[pre] (0,-1.2) rectangle ++(1,0.8); \node[anchor=west] at (1.1,-0.8) {prelude/coda};
  \draw[live] (7,-1.2) rectangle ++(1,0.8); \node[anchor=west] at (8.1,-0.8) {recurrent};
  \draw[dead] (13.5,-1.2) rectangle ++(1,0.8); \node[anchor=west] at (14.6,-0.8) {dead slot};
\end{tikzpicture}
\caption{KV layout of one Huginn token page. Upstream reserves $8\times32=256$ layer slots, of which 124 are
dead. Packing stores the 132 live slots in 17 blocks of eight instead of 32.}
\label{fig:packing}
\end{figure}

Packing gives each depth plane only the layer slots that it can hold (\cref{fig:packing}). A KV page becomes a
run of blocks, each with \texttt{num\_layers} slots, and slot $s$ of block $b$ is entry
$b\cdot\texttt{num\_layers}+s$ of a flattened view of the pool. Plane 0 maps all eight layers; planes 1--31 map
the four recurrent layers. A Huginn page then needs 17 blocks instead of 32. Attention, KV writes, early-exit
plane copies and the asynchronous finalize kernel take an entry stride and a first-layer offset instead of
assuming a full plane. The PD transport sends each plane's live slot range, and the PD coordinator prices
requests in blocks per page. Values never change, only their addresses, so outputs are bitwise identical. For
models without boundary layers, such as Ouro, the layout is unchanged. These checks assume equal inputs and
equal batch composition. When packing admits more requests, the scheduler also changes the batches; we report
those token comparisons separately in \cref{sec:hugcomp}.

\subsection{Rounding fences make fused kernels bitwise (\rung{0})}
\label{sec:fusion}

PyTorch computes a bf16 RMSNorm as a chain of eight kernels: cast to fp32, square, mean, add $\epsilon$, rsqrt,
multiply, cast to bf16, and multiply by the weight, plus an addition when a residual is fused. A one-kernel
Triton version differs from it in two ways. It sums the squares in a different order, and its compiler moves
rounding points. We keep both under control.

\paragraph{Keep PyTorch's summation order.} The exact RMSNorm sums the squares in PyTorch's order. Its first form
leaves the mean to PyTorch between a Triton square kernel and a fenced Triton tail for $\mathrm{rsqrt}(\bar{x}+\epsilon)$
and the scaled output (Triton's \texttt{rsqrt} lowers to PyTorch's instruction; $\bar{x}+\epsilon$ is one IEEE addition).
PyTorch~2.14's mean fixes its order by its launch configuration ($W$ lanes per row set by the row count, four
strided accumulators per lane, a halving tree); for Ouro's and Huginn's widths the second form repeats those
additions in one kernel, and shapes PyTorch would split across warps keep three kernels.

\paragraph{Fence the roundings.} Writing the same casts as PyTorch is not enough. Triton and LLVM keep an fp32
product live across a cast to bf16 and back (\texttt{.to(bf16).to(fp32)}), and contract $a\cdot b+c$ into a fused multiply-add. Our first
fused residual epilogue therefore computed $\mathrm{round}(r+w\cdot n)$ while PyTorch computes
$\mathrm{round}(r+\mathrm{round}(w\cdot n))$, a double rounding~\cite{higham2002accuracy}. On \DiagN{} bf16
elements, \DiagDouble\% differed from PyTorch by one ulp, and \DiagSingle{} differed from a single-rounding oracle.
A \emph{rounding fence} is an inline PTX instruction, \texttt{cvt.rn.bf16.f32} or \texttt{mul.rn.f32}, that
performs one IEEE rounding the compiler cannot see through, so it can neither drop nor merge it. The exact RoPE
kernel fences each of its two products before adding them. The exact SiLU-and-multiply kernel fences
$\mathrm{silu}(g)$ before multiplying by the up projection; its exponential comes from the same CUDA math
library that PyTorch links. For the residual addition we use PyTorch's own addition.

\paragraph{Fuse only inside graphs.} On the GH200's Grace host a Triton launch from Python costs more host time
than the PyTorch launches it replaces. Under CUDA graphs that cost is paid once at capture; in eager prefill it is
paid on every call. ExactLoop therefore uses the fused kernels only while a CUDA graph is being captured, which
covers decode, and leaves eager code paths on PyTorch.

\subsection{Loop fusion changes only which rows share a batch (\rung{1})}
\label{sec:loops}

Rows decoding at a fixed depth run all their loops in one submission, which replays one graph for $L$ loops.
On the GH200 this cut Ouro's submissions per token from \SubsOuroMain{} to \SubsOuroOpt{} and Huginn's from
\HugSubsMain{} to \HugSubsOpt{} at 8 requests. Each row's arithmetic is unchanged: with one request at a time,
GSM8K token sequences matched on all \GsmN{} questions. The batches are not unchanged. The base engine staggers
rows across loop phases, while loop fusion runs them together, so GEMMs see different batch sizes, cuBLAS may pick
different algorithms, and some bf16 outputs change at high
load: \IsoMainOptSame{} requests matched the base engine at 8 to 64 requests, while two runs of the base engine
matched on \IsoMainSelfSame. That makes it \rung{1}.

A fused step is $L$ loops long, and the scheduler admitted a fixed number of prefill batches per decode step. Prefill was
therefore admitted $L$ times less often per unit of decode work, and TTFT grew under load (\cref{sec:sessions}).
ExactLoop admits up to $L$ prefill batches after an $L$-loop fused step, which restores the base engine's ratio of
prefill to decode work per loop.

\subsection{Per-plane coverage makes prefixes reusable; position keys fix the noise order (\rung{2})}
\label{sec:noise}

We changed the publication rule so that each plane must cover only the layers that it holds. Huginn prefixes are
then published and reused, but a cache hit still changes outputs. Huginn draws the initial latent state of each
token from a per-request random stream, in the order tokens are processed. A request whose first $k$ prompt
tokens come from the cache draws $k$ fewer values, so every later token receives different noise than in a cold
run. The prefill chunk size changes the draws in the same way.

Position-keyed noise makes the noise a function of the request seed, the token position and the channel. A
hash in the style of SplitMix64~\cite{steele2014splitmix}, applied in stages with distinct constants, turns
(seed, position, channel) into a 24-bit uniform number, and an inverse normal CDF maps it onto the intended
truncated normal distribution at 24-bit resolution, as in counter-based generators~\cite{salmon2011random123}.
The target distribution is the same, subject to this discretization and the inverse-CDF implementation's rounding.
64-bit wrap-around is identical on CPU and GPU. In CPU tests, cached and cold runs, and prefill chunks of 2 and 8
tokens, produced identical tokens. A seeded run sees different samples from upstream's sequential generator, so
the feature is opt-in together with Huginn prefix caching.

On the GPU the noise is only part of the story. We ran \CacheNoiseN{} prompts of 512 tokens one at a time on
Huginn-0125, cold and then again with \CacheHitTokens{} tokens served from the cache; cold runs repeated exactly
(\CacheColdRepeatSame). With position-keyed noise, \CacheNoiseFullSame{} cached runs matched the cold run token
for token (95\% interval \CacheNoiseFullLo--\CacheNoiseFullHi\%), and a longer prompt that reused the same prefix
matched in \CacheNoisePartSame{} (\CacheNoisePartLo--\CacheNoisePartHi\%). An earlier eight-prompt check with the
sequential generator gave \CacheSeqFullSame{} full hits; there, position keys fixed the one prompt that changed at
its first token. The remaining divergences come from prefill arithmetic: a cached run prefills only the suffix, so
its GEMMs see fewer rows, and changing only the prefill chunk size also changed outputs (\CacheChunkSame{} cold
runs matched between chunks of 2048 and 128 tokens). On the GPU, prefix reuse therefore behaves like a change in
batch composition (\rung{1}) layered on a \rung{2} change of samples. The noise itself is not negligible at 32
loops: the two generators gave the same tokens for only \CacheNoiseKindSame{} cold runs. Position keying removes
one source of cache dependence without making all GPU cache hits token-identical.

\subsection{Packed prefill and its accuracy check (\rung{3})}
\label{sec:packedprefill}

The engine resolves packed prefill at construction: it turns it on for CUDA, the LAST\_EXITED layout, refill
scheduling and a backend with a packed path, and an explicit setting always wins. Packed bf16 prefill computes
attention with different arithmetic than per-row prefill, so a few greedy sequences change (\cref{tab:ladder}).
For Huginn, whose boundary layers need their own attention metadata, packed prefill changed the full GSM8K split by
\FullHugPackedDelta{} points (\cref{sec:gates}), within the one-point rule only on its point estimate. The packed path stages its inputs in host-mapped pinned memory (UVA) instead of copying them to the GPU.

\subsection{The PD fixes change no values (\rung{0})}
\label{sec:pdfixes}

The upstream PD coordinator built every model with Ouro's configuration class, so a real Huginn checkpoint
could not start; it now picks the class from the model type the worker reports. With packing, the transport
sends only live slots, \PdHugRatio{} of the full Huginn layout. Descriptors that are contiguous on both sides are
coalesced, and an overlap counter now credits transfers issued during a prefill computation. The connector can
also create its UCX~\cite{shamis2015ucx} backend explicitly with an error-handling mode, and the default transfer
budget now holds one long prompt's KV in flight (\cref{sec:lessons}).

\section{Evaluation}
\label{sec:eval}

\subsection{Setup}
\label{sec:setup}

\paragraph{Platforms.} We used three machines. A batch-scheduled GH200 system ran single-GPU jobs and the
two-GPU PD jobs, with PyTorch~2.14, Triton~3.8 and NIXL~1.4.1. A standalone GH200 host, whose GPU our runs held
exclusively, ran the release table, the accuracy gates and the end-to-end and Huginn session comparisons. A rented RTX~5090 ran the first
characterization. Every speedup compares arms run on the same machine.

\paragraph{Workloads.} In a burst workload, $C$ requests arrive at once, each with a prompt of 512 or 2048
random tokens and a fixed number of greedy output tokens. Session workloads replay multi-turn traces of a
coding agent~\cite{semianalysis2026cctraces} whose prompts
grow with the conversation, with $n$ concurrent sessions and the prefix cache on. Ouro runs at depth four with
asynchronous scheduling and CUDA graphs unless noted; Huginn runs at depth 32 with CUDA graphs and synchronous
scheduling, which the engine requires for it.

\paragraph{Configurations.} \Cref{tab:release} measures the released defaults (loop fusion with the prefill
share, plane packing, exact kernels in graphs and in eager calls over 256 rows, automatic packed prefill) against
the base engine, three runs per cell. ExactLoop evolved during the study, so the component tables that follow state which features their arms had;
unless they say otherwise, their ExactLoop arm prefilled one row at a time and had no prefill share.

\paragraph{Noise.} Apart from \cref{tab:release} and the arms that say so, cells are single runs. On the standalone host two runs of the base engine at 32 requests
differed by a factor of \ScopeRepeatThirtyTwo, and the first repeat of an arm ran \FirstRepSlow\% below its later
repeats (up to \SessFirstSlowMax\% for session arms). Where an arm has repeats we drop the first; we treat differences within the rerun spread as ties. Several gains reported below, such as the graph-count and packing ratios, lie within or close to this band and are presented as directional. Token
identity compares each request with the request of the same level, repeat and index.

\paragraph{Run-to-run variation.} The headline cells retain two runs after warm-up, while several component
experiments have only one. Host and scheduling variation can therefore obscure small differences. We discuss
component effects only when they exceed the observed rerun spread and leave the remaining measurements as
directional. \cref{sec:gates} treats the accuracy subset separately because its sample size differs from the
performance runs.

\subsection{The released defaults are \RelAllMin--\RelAllMax\X{} faster than the base engine}
\label{sec:release}

\begin{table}[t]
\caption{Released defaults against the base engine, standalone GH200, burst workloads with 512- or 2048-token
prompts (Ouro: 128 or 64 output tokens, async scheduling and graphs; Huginn: 64 output tokens, 32 loops, sync and
graphs), 64 graphs, \RelFallbacks{} eager fallbacks. Output tokens/s; three runs per cell, first dropped, mean of \RelN.
All rows come from one job and one build, the paper tree; each ratio is paired within the job.}
\label{tab:release}
\centering\small\setlength{\tabcolsep}{2pt}
\resizebox{\columnwidth}{!}{
\begin{tabular}{rrrrrrr}
\toprule
model, prompt & $C$ & base & ExactLoop & speedup & TTFT p50 b/E (s) & TPOT p50 b/E (ms) \\
\midrule
Ouro, 512 & 1 & 55.9 & 109.2 & 1.95$\times$ & 0.19/0.09 & 16.6/8.5 \\
Ouro, 512 & 8 & 339.9 & 657.4 & 1.93$\times$ & 0.29/0.14 & 19.5/10.2 \\
Ouro, 512 & 32 & 677.9 & 1454.2 & 2.15$\times$ & 1.05/0.51 & 37.1/17.2 \\
Ouro, 512 & 64 & 901.2 & 1483.4 & 1.65$\times$ & 2.30/1.28 & 50.8/31.7 \\
Ouro, 2048 & 8 & 102.4 & 208.8 & 2.04$\times$ & 1.66/0.81 & 46.2/22.9 \\
Ouro, 2048 & 32 & 129.3 & 246.9 & 1.91$\times$ & 6.60/3.30 & 137.9/75.1 \\
Huginn, 512 & 1 & 11.3 & 19.9 & 1.77$\times$ & 0.30/0.25 & 85.2/46.9 \\
Huginn, 512 & 8 & 48.3 & 109.3 & 2.26$\times$ & 1.23/0.43 & 129.7/58.5 \\
Huginn, 512 & 32 & 72.7 & 206.6 & 2.84$\times$ & 4.72/2.02 & 227.9/118.2 \\
Huginn, 512 & 64 & 65.0 & 205.3 & 3.16$\times$ & 22.66/3.95 & 270.1/121.0 \\
\bottomrule
\end{tabular}
}
\end{table}

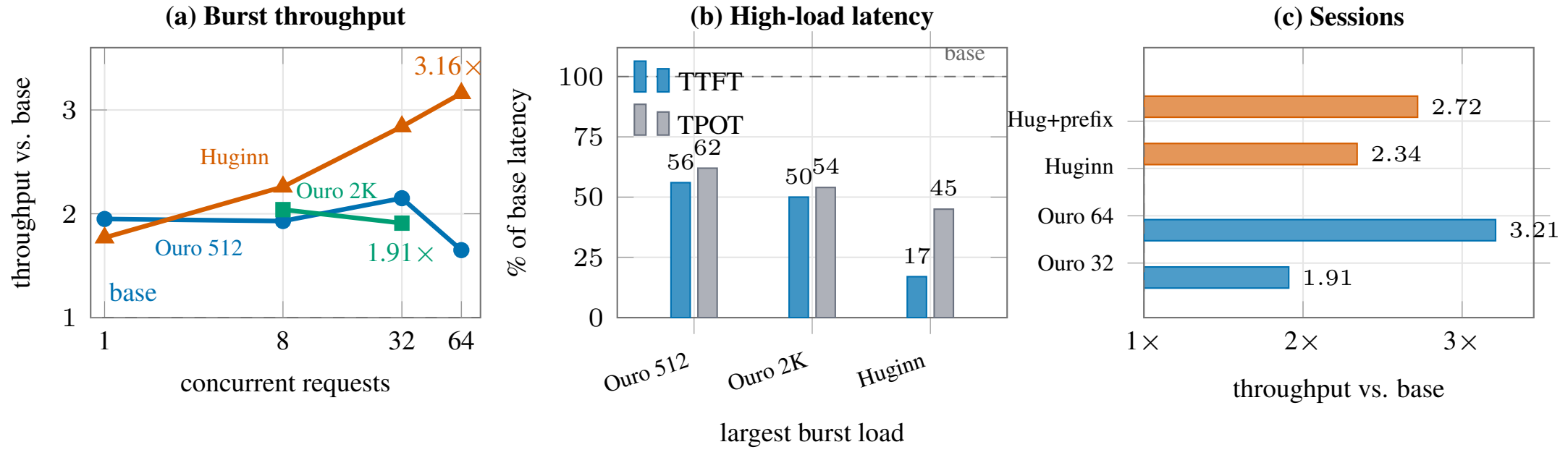
\begin{figure*}[t]
\centering
\definecolor{exactblue}{HTML}{0072B2}
\definecolor{exactteal}{HTML}{009E73}
\definecolor{exactorange}{HTML}{D55E00}
\definecolor{exactgray}{HTML}{6B7280}

\begin{tikzpicture}
\begin{groupplot}[
  group style={group size=3 by 1, horizontal sep=1.25cm},
  width=0.30\textwidth,
  height=4.05cm,
  tick label style={font=\scriptsize},
  label style={font=\scriptsize},
  title style={font=\scriptsize\bfseries, yshift=-1.1ex},
  legend style={font=\scriptsize, draw=none, fill=none, inner sep=1pt},
  axis line style={draw=black!55},
  tick style={draw=black!45},
  grid=major,
  grid style={draw=black!10},
]

\nextgroupplot[
  title={(a) Burst throughput},
  xmode=log,
  log basis x=2,
  xtick={1,8,32,64},
  xticklabels={1,8,32,64},
  xmin=0.85,
  xmax=80,
  ymin=1,
  ymax=3.6,
  ytick={1,2,3},
  xlabel={concurrent requests},
  ylabel={throughput vs. base},
]
\addplot[exactblue, very thick, mark=*, mark size=1.5pt]
  coordinates {(1,1.95) (8,1.93) (32,2.15) (64,1.65)};
\addplot[exactteal, very thick, mark=square*, mark size=1.4pt]
  coordinates {(8,2.04) (32,1.91)};
\addplot[exactorange, very thick, mark=triangle*, mark size=1.8pt]
  coordinates {(1,1.77) (8,2.26) (32,2.84) (64,3.16)};
\draw[black!45, dashed] (axis cs:0.85,1) -- (axis cs:80,1);
\node[anchor=south west, font=\scriptsize, text=exactblue] at (axis cs:0.9,1.03) {base};
\node[anchor=north, font=\tiny, text=exactblue] at (axis cs:3,1.88) {Ouro 512};
\node[anchor=south, font=\tiny, text=exactteal] at (axis cs:15,2.02) {Ouro 2K};
\node[anchor=south east, font=\tiny, text=exactorange] at (axis cs:8,2.3) {Huginn};
\node[anchor=north, font=\scriptsize, text=exactteal] at (axis cs:32,1.85) {1.91$\times$};
\node[anchor=south, font=\scriptsize, text=exactorange] at (axis cs:56,3.18) {3.16$\times$};

\nextgroupplot[
  title={(b) High-load latency},
  ybar,
  bar width=5pt,
  xmin=0.35,
  xmax=3.65,
  ymin=0,
  ymax=112,
  xtick={1,2,3},
  xticklabels={Ouro 512,Ouro 2K,Huginn},
  xticklabel style={font=\tiny, rotate=18, anchor=north east},
  xlabel={largest burst load},
  ylabel={\% of base latency},
  ytick={0,25,50,75,100},
  legend pos=north west,
  legend columns=1,
  nodes near coords,
  nodes near coords style={font=\tiny, /pgf/number format/fixed, /pgf/number format/precision=0},
]
\addplot[fill=exactblue!75, draw=exactblue] coordinates {(1,56) (2,50) (3,17)};
\addlegendentry{TTFT}
\addplot[fill=exactgray!55, draw=exactgray] coordinates {(1,62) (2,54) (3,45)};
\addlegendentry{TPOT}
\draw[black!55, densely dashed] (axis cs:0.35,100) -- (axis cs:3.65,100);
\node[anchor=south east, font=\tiny, text=black!60] at (axis cs:3.58,101) {base};

\nextgroupplot[
  title={(c) Sessions},
  xbar,
  bar width=5.5pt,
  xmin=1,
  xmax=3.45,
  ymin=0.4,
  ymax=5.0,
  xtick={1,2,3},
  xticklabels={1$\times$,2$\times$,3$\times$},
  ytick={1,2,3,4},
  yticklabels={Ouro 32,Ouro 64,Huginn,Hug+prefix},
  yticklabel style={font=\tiny},
  enlarge y limits=0.12,
  xlabel={throughput vs. base},
  nodes near coords,
  nodes near coords style={font=\tiny, /pgf/number format/fixed, /pgf/number format/precision=2},
]
\addplot[fill=exactblue!70, draw=exactblue] coordinates {
  (\SsFinalGainThirtyTwo,1) (\SsFinalGainSixtyFour,2)
};
\addplot[fill=exactorange!65, draw=exactorange] coordinates {
  (\HugSessGainExact,3) (\HugSessGain,4)
};
\draw[black!55, densely dashed] (axis cs:1,0.4) -- (axis cs:1,4.6);
\end{groupplot}
\end{tikzpicture}
\caption{Main serving results for the released defaults. (a) Throughput speedup over the base engine as concurrency
increases; the dashed line marks no speedup. (b) TTFT and TPOT at the largest tested burst load, as a percentage of
base latency; lower is better. (c) Throughput speedup on multi-turn agent sessions, including Huginn with and without
prefix caching. Values are derived from \cref{tab:release,tab:hugsess} (standalone GH200) and, for Ouro sessions,
\appref{tab:sessions} (batch-scheduled GH200); markers and bars use the mean of the retained runs.}
\label{fig:release}
\end{figure*}

\Cref{tab:release} is the headline result. \Cref{fig:release} plots it. With 512-token prompts the released defaults serve Ouro
\RelOuroShortMin--\RelOuroShortMax\X{} and Huginn \RelHugMin--\RelHugMax\X{} faster than the base engine, and both
time to first token and time per output token fall at every load. With 2048-token prompts, where packed prefill
dominates, Ouro runs \RelOuroLongMin--\RelOuroLongMax\X{} faster. Huginn's median TTFT at 64 requests falls from
\RelHugTtftMainSixtyFour{} to \RelHugTtftOptSixtyFour~s. Packed prefill and batch composition change some greedy
sequences: \RelOuroShortSame{}, \RelOuroLongSame{} and \RelHugSame{} requests are identical to the base engine.

On the batch-scheduled GH200, with Ouro and Huginn on different builds, the same workloads gave the same token
identity but larger Ouro gains (\RelBOuroShortMin--\RelBOuroShortMax\X{} and \RelBOuroLongMin--\RelBOuroLongMax\X) and
Huginn gains of \RelBHugMin--\RelBHugMax\X: the speedups depend on the platform. The base engine's batch-1 Huginn throughput also moved by \HugBaseDriftOne\% between two batches, so
the batch-1 Huginn ratio is the least certain of these headline numbers.

On the batch-scheduled host, in four interleaved rounds per arm, the released defaults served Ouro agent sessions
\SsFinalGainThirtyTwo\X{} and \SsFinalGainSixtyFour\X{} faster than the base engine at 32 and 64 sessions, with a
median TTFT of \SsFinalTtftRelSixtyFour{} instead of \SsFinalTtftBaseSixtyFour~s at 64, and Huginn sessions
\HugSessGain\X{} faster with the prefix cache (\cref{tab:hugsess}); \appref{tab:sessions} takes the Ouro gain apart.

\subsection{Real prompts under Poisson arrivals}
\label{sec:online}

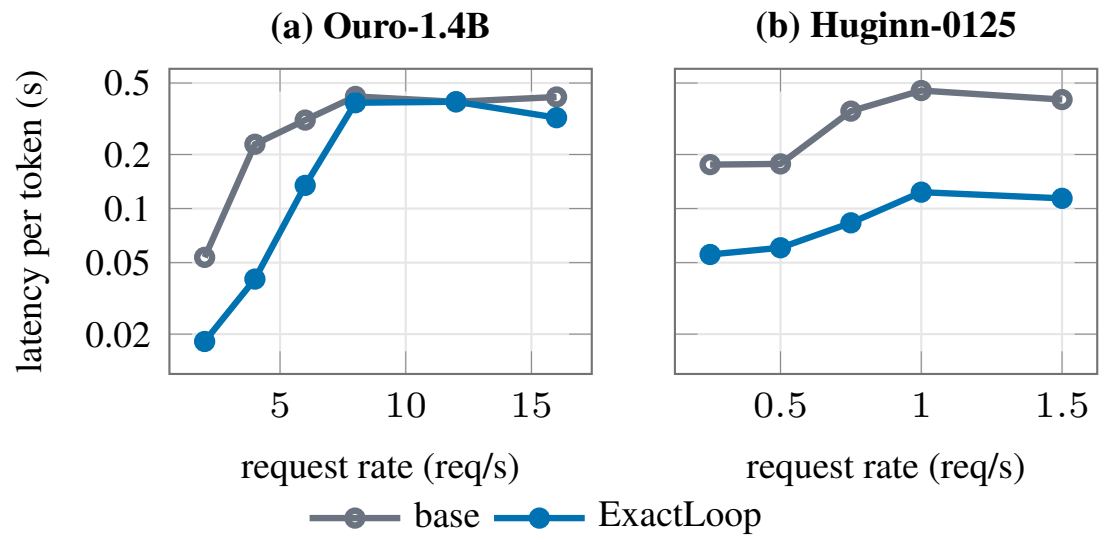
\begin{figure}[t]
\centering
\definecolor{exactblue}{HTML}{0072B2}
\definecolor{exactorange}{HTML}{D55E00}
\definecolor{exactgray}{HTML}{6B7280}
\begin{tikzpicture}
\begin{groupplot}[
  group style={group size=2 by 1, horizontal sep=0.55cm},
  width=0.53\columnwidth,
  height=3.6cm,
  tick label style={font=\scriptsize},
  label style={font=\scriptsize},
  title style={font=\scriptsize\bfseries, yshift=-1.1ex},
  legend style={font=\scriptsize, draw=none, fill=none, inner sep=1pt},
  axis line style={draw=black!55},
  tick style={draw=black!45},
  grid=major,
  grid style={draw=black!10},
  ymode=log,
  ymin=0.012, ymax=0.6,
  ytick={0.02,0.05,0.1,0.2,0.5},
  yticklabels={0.02,0.05,0.1,0.2,0.5},
  yminorticks=false,
  xlabel={request rate (req/s)},
]
\nextgroupplot[title={(a) Ouro-1.4B}, ylabel={latency per token (s)}, legend to name=onlinelegend, legend columns=2]
\addplot[exactgray, very thick, mark=o, mark size=1.4pt] table[x=rate, y=base] {figs/online-ouro.dat};
\addlegendentry{base}
\addplot[exactblue, very thick, mark=*, mark size=1.4pt] table[x=rate, y=opt] {figs/online-ouro.dat};
\addlegendentry{ExactLoop}
\nextgroupplot[title={(b) Huginn-0125}, yticklabels={}]
\addplot[exactgray, very thick, mark=o, mark size=1.4pt] table[x=rate, y=base] {figs/online-hug.dat};
\addplot[exactblue, very thick, mark=*, mark size=1.4pt] table[x=rate, y=opt] {figs/online-hug.dat};
\end{groupplot}
\end{tikzpicture}
\\[-1ex]\ref*{onlinelegend}
\caption{ShareGPT prompts with their reference answer lengths (prompt ${\le}1024$, answer ${\le}256$ tokens), Poisson
arrivals, 256 (Ouro) or 80 (Huginn) requests per rate, standalone GH200, last of two runs; median end-to-end latency
per output token (log scale; lower is better).}
\label{fig:online}
\end{figure}

We replayed ShareGPT and Alpaca prompts with each reference answer's length, offline and under Poisson arrivals,
on the standalone host and the same two trees (\cref{fig:online}). Offline, with 64 requests at once, they were \OnOuroOffShareGPTGain\X{}
(ShareGPT) and \OnOuroOffAlpacaGain\X{} (Alpaca) faster for Ouro and \OnHugOffShareGPTGain\X{} and
\OnHugOffAlpacaGain\X{} for Huginn; Alpaca's answers average \OnAlpacaOutMean{} tokens, so admission weighs
more. Online, Huginn's latency per token was \OnHugNlGainMin--\OnHugNlGainMax\X{} lower at every rate, and its
ExactLoop arm had not saturated at \OnHugRateMax{} requests/s. Ouro's base engine saturated first: its latency per
token rose \OnOuroBaseKnee\X{} from 2 to 4 requests/s, while ExactLoop stayed at \OnOuroOptFour~s at 4 (base:
\OnOuroBaseTwo~s at 2) and was \OnOuroNlSix\X{} lower at 6. From \OnOuroSatRate{} requests/s both were saturated
at the same throughput (\OnOuroPlateauBaseLo--\OnOuroPlateauBaseHi{} against
\OnOuroPlateauOptLo--\OnOuroPlateauOptHi{} tokens/s), so that ceiling lies in what ExactLoop leaves unchanged; we
did not trace the step mix. Median TTFT was higher with ExactLoop at \OnOuroTtftWorseN{} of \OnOuroRateN{} Ouro
and \OnHugTtftWorseN{} of \OnHugRateN{} Huginn rates. Ouro-2.6B (48 layers, same shapes) ran \OuroBigShortMin--\OuroBigShortMax\X{} and
\OuroBigLongMin--\OuroBigLongMax\X{} faster on the release workloads (\OuroBigShortSame{} and \OuroBigLongSame{}
identical) and \OuroBigOffGain\X{} on ShareGPT. \OnAbortRuns{} of \OnPlainRuns{} Poisson runs of Ouro's released
defaults aborted, in a cuBLAS GEMM writing to an unmapped address; the cause is open.

\subsection{Packing lifts Huginn's ceiling; loop fusion carries batch 1}
\label{sec:hugcomp}

\Appref{tab:packing} isolates packing on an earlier tree. The same pool holds \PackUnpackedTokens{} tokens unpacked and
\PackPackedTokens{} packed. At 8 requests both layouts fit and run alike (\PfGainEight\X); at 32 and 64 requests
packing raised throughput \PfGainThirtyTwo\X{} and \PfGainSixtyFour\X, and at 64 requests it cut the median TTFT from
\PfTtftUnpSixtyFour{} to \PfTtftPackSixtyFour~s. \PfSame{} requests matched: packing changes no value, but when more
requests fit, the batches differ, as with loop fusion. In an earlier tree with per-row prefill, where the batches
did not change, all \PackSame{} requests matched. The first GH200 batch measured smaller gains of \PackGainThirtyTwo\X{} and
\PackGainSixtyFour\X{} at 32 and 64 requests (ledger item 10). We report the repeated earlier-tree measurement above,
and the lower values lie within the rerun spread.

At one request capacity does not bind, so the batch-1 gain of \cref{tab:release} comes from the other changes. In one
batch on that tree, the fused kernels contributed only \HsplitKernOne\X, \HsplitKernEight\X{} and
\HsplitKernThirtyTwo\X{} at 1, 8 and 32 requests (full set over kernels disabled, three runs each, first dropped):
Huginn's layers are large and few, so kernel count matters little, and at one request the remaining changes are loop
fusion and the host path. Against a base-engine arm from a later batch the full gain was
\HsplitFullOne\X, \HsplitFullEight\X{} and \HsplitFullThirtyTwo\X{} on the standalone host (\RelHugOne\X{} at one request
in \cref{tab:release}); the base engine's batch-1 Huginn speed moved by \HugBaseDriftOne\% between two batches. Its 32 loops make submissions matter more than for Ouro: at 8 requests
the base engine needed \HugSubsMain{} submissions per output token and ExactLoop \HugSubsOpt. Packed prefill helps
Huginn as much as Ouro: at 32 requests it was \HugPackedPfiveGainThirtyTwo\X{} faster with 512-token prompts and
\HugPackedPtwokGainThirtyTwo\X{} with 2048-token prompts, where it cut the median TTFT from \HugPackedPtwokTtftRows{}
to \HugPackedPtwokTtftPacked~s (single runs).

\subsection{Prefix caching makes Huginn sessions \HugSessGainNopc\X{} faster on top of ExactLoop}
\label{sec:hugsess}

\begin{table}[t]
\caption{Huginn-0125 agent sessions, 16 sessions, at most four turns and 2000-token prompts, \HugSessReqs{}
requests, standalone GH200, paper tree, 64 graphs. ExactLoop: released defaults with position-keyed noise; ``+ prefix cache''
adds Huginn prefix caching. Four interleaved rounds per arm, the first dropped, mean of three.}
\label{tab:hugsess}
\centering\small\setlength{\tabcolsep}{3pt}
\resizebox{\columnwidth}{!}{
\begin{tabular}{@{}lrrrrr@{}}
\toprule
arm & tok/s & TTFT p50 & p90 (s) & TPOT (s) & hit \% \\
\midrule
base & 12.2 & 5.52 & 11.94 & 0.64 & -- \\
ExactLoop & 28.6 & 0.48 & 0.96 & 0.37 & 0 \\
+ prefix cache & 33.3 & 0.36 & 0.69 & 0.27 & 78 \\
\bottomrule
\end{tabular}
}
\end{table}

Prefix caching with position-keyed noise reused \HugSessPcHit\% of prompt tokens and raised Huginn session throughput
\HugSessGainNopc\X{} over ExactLoop without the cache, which was itself \HugSessGainExact\X{} the base engine
(\cref{tab:hugsess}); together \HugSessGain\X, with the median TTFT cut to 1/\HugSessTtftRatio. The base engine ran
alike with 16 and 64 graphs (\HugSessBaseSixtyFour{} tokens/s), and no Huginn session run fell back to eager
execution. On an earlier tree the prefill share made no measurable difference here. On the batch-scheduled host an
earlier tree gave \HugSessGainB\X{} with the prefix cache.

\subsection{Exact fusion matches approximate throughput}
\label{sec:kernels}

With the three-kernel RMSNorm the exact set cut kernels per token from \KernOffPerToken{} to \KernExactPerToken{} at
batch 1, a \KernExactCut\% reduction, and GPU time per token from \KernOffGpuMs{} to \KernExactGpuMs~ms. With the one-kernel RMSNorm it
raised throughput \KfExactOne\X, \KfExactEight\X, \KfExactThirtyTwo\X{} and \KfExactOneTwoEight\X{} at 1, 8, 32 and
128 requests, against \KfThreeMin--\KfThreeMax\X{} with the three-kernel norm (\appref{tab:kernels}), and
\KfExactSame{} requests matched the unfused arm. The approximate set, whose RMSNorm sums in its own order and which
has no fences, was no faster within the host's spread (\KfApproxMin--\KfApproxMax\X) and left only
\KfApproxSame{} requests unchanged. ncu on the three-kernel form shows why kernel count, not kernel work, sets the
batch-1 time: its four exact kernels ran \NcuMinUs--\NcuMaxUs~$\mu$s each and reached at most \NcuDramMax\% of DRAM
throughput. The exact set therefore costs no measurable throughput in this experiment. We discuss its accuracy
result in \cref{sec:gates}.

\Appref{tab:scope} shows why fusion was first limited to graphs: fused kernels in every scope made eager prefill
slower than the base engine's at all three levels, and graph-only fusion made it faster. Its throughput columns are
single runs within the rerun spread, so we use only its prefill times. Packed prefill
makes eager calls large, though: the exact kernels in eager calls over 256 token rows cut Ouro's prefill GPU time by
\RowsOuroPrefillCut\% and Huginn's prefill time by \RowsHugPrefillCut\% at 64 requests and raised throughput up to
\RowsOuroLongHi\X{} (Ouro, 2048-token prompts) and \RowsHugHi\X{} (Huginn). Fusing Huginn's own activation and
rotary embedding added \HugFuseLo--\HugFuseHi\X, and the one-kernel RMSNorm, \OneBenchLo--\OneBenchHi\X{} faster
per norm in CUDA graphs and bitwise in all \OneCases{} GPU test cases, added \OneOuroShortLo--\OneOuroLongHi\X{}
for Ouro and \OneHugLo--\OneHugHi\X{} for Huginn. Running Ouro's q/k/v and gate/up projections
as one GEMM each over concatenated weights (seven GEMMs per layer and loop to four) is bitwise on
this cuBLAS for each of rows 1--\ProjSweepTop, eagerly and in graph replay (not on an L40S, so only checked GPUs fuse); it cut decode GPU time at one request by \ProjDecodeCut\%
and added \ProjShortLo--\ProjShortHi\X{} (512-token prompts) and \ProjLongLo--\ProjLongHi\X{} (2048).
No request changed in these arms; \cref{tab:release} includes them.

\subsection{The prefill share trades time per token for time to first token}
\label{sec:sessions}

\Appref{tab:sessions} shows a trade-off. With 64 sessions, ExactLoop without the prefill share and with per-row
prefill cut the median TPOT from \SsBaseTpotSixtyFour{} to \SsRowsTpotSixtyFour~s but raised the median TTFT from
\SsBaseTtftSixtyFour{} to \SsRowsTtftSixtyFour~s, because fused decode steps delayed prefill admission
(\cref{sec:loops}); we compare against the 64-graph base engine, although its median TTFT was higher than with 16
graphs (\SsBaseSixteenTtftSixtyFour~s), which we cannot explain. Packed prefill raised throughput to \SsPackedGainSixtyFour\X{} the base engine and brought the
median TTFT down to \SsPackedTtftSixtyFour~s. With 2048-token bursts,
packed prefill was \PackedPtwokEight\X{} and \PackedPtwokThirtyTwo\X{} faster; with 512-token bursts it was within
noise of per-row prefill (\UvaPackedEight\X{} and \UvaPackedThirtyTwo\X{} at 8 and 32 requests, against a rerun of
\UvaRepeatEight\X{} and \UvaRepeatThirtyTwo\X), and without UVA staging it fell to \UvaNouvaEight\X{} and
\UvaNouvaThirtyTwo\X{} (\UvaN{} kept runs each).

The prefill share helps the burst case, and was ahead in both kept runs. At 32 requests with 512-token prompts
(\ShareN{} kept runs), it raised throughput from \BurstNoshareTput{} to \BurstShareTput{} tokens/s, cut the median TTFT from
\BurstNoshareTtftFifty{} to \BurstShareTtftFifty~s and the p90 from \BurstNoshareTtftNinety{} to
\BurstShareTtftNinety~s, and lowered TPOT from \BurstNoshareTpot{} to \BurstShareTpot~ms. In 64-session runs it is
a trade-off: on top of packed prefill, the share cut the median TTFT from \SsPackedTtftSixtyFour{} to
\SsRelTtftSixtyFour~s and the p90 from \SsPackedTtftNinetySixtyFour{} to \SsRelTtftNinetySixtyFour~s, but raised the
median TPOT from \SsPackedTpotSixtyFour{} to \SsRelTpotSixtyFour~s and kept \SsShareGainSixtyFour\X{} of the
throughput; with 32 sessions it changed little (\SsShareGainThirtyTwo\X). We
made the share the default because TTFT is the metric agents wait on, but the choice depends on the service
objective.

\subsection{Accuracy checks and the approximate-kernel decision}
\label{sec:gates}

We used the engine's GSM8K~\cite{cobbe2021gsm8k} harness on a fixed \GsmN-question subset, greedy, one request
at a time, with a regression limit of one point. On the GH200, the exact kernel set answered \GateOuroExact{}
questions correctly, as did the unfused engine, and \GateOuroSeqDiff{} token sequences differed. On the
RTX~5090, loop fusion matched the base engine on all sequences and exit depths (\GsmNoneCand{} correct). The
approximate RMSNorm alone scored \GsmNormCand{} against \GsmNormRef{} (\GsmNormDelta{} points, paired standard
error \GsmNormSe; \GsmNormSeqDiff{} sequences differed) and the full approximate set \GsmAllCand{}
(\GsmAllDelta{}, standard error \GsmAllSe; \GsmAllSeqDiff{} differed). Both trip our one-point rule. We used that
rule to decide whether to ship the kernels, not as a hypothesis test; the 95\% intervals, roughly $[-5.5,+0.9]$
and $[-8.5,+1.6]$, include zero. With \GsmN{} questions, the subset cannot separate changes of one and two
points, so we repeated both decisions on all \FullN{} test questions on a GH200, batched (32 requests, CUDA graphs,
same stops and scorer). The exact and unfused paths then matched on every sequence. The approximate
set changed the score by \FullOuroApproxDelta{} points (s.e.\ \FullOuroApproxSe) and the approximate RMSNorm by
\FullOuroApproxNormDelta{} (\FullOuroApproxNormSe), within the rule; GPU, batching and reference differ from the
subset. The approximate kernels stay unshipped: they are neither bitwise nor faster.

Packed prefill changed Ouro by \FullOuroPackedDelta{} points (s.e.\ \FullOuroPackedSe) and Huginn by
\FullHugPackedDelta{} (\FullHugPackedSe):
within the rule on the point estimate, with 95\% intervals reaching \FullOuroPackedLo{} and \FullHugPackedLo.
Batch composition alone moves scores as much: with 16 requests per batch, per-row Ouro answered
\FullOuroRowsSixteen{} instead of \FullOuroRows{} and packed prefill changed it by \FullOuroPackedDeltaBSixteen.
Batches of 64 replayed the 32-request outputs almost token for token, and Huginn's batch 16 replayed them exactly
(\FullHugSixteenSame{} of \FullN), so Huginn has one observation and a one-point loss is not excluded.
On MATH-500 (\MathN{} questions) packed prefill changed Ouro by \MathOuroPackedDelta{} (\MathOuroPackedSe) and
Huginn by \MathHugPackedDelta{} points (\MathHugPackedSe), and the exact set matched unfused on all \MathN.

Huginn's \rung{2} changes were gated with all \GateSmallN{} questions run as one batch on the released defaults.
Sequential noise answered \GateSmallSeq{} correctly, position-keyed noise \GateSmallPos{} (\GateSmallSeqPosDelta{}
points, paired standard error \GateSmallSeqPosSe), and position-keyed noise with prefix reuse \GateSmallCache{}
(\GateSmallSeqCacheDelta{} against sequential noise, standard error \GateSmallSeqCacheSe). Both miss the same
one-point rule, although their differences are within about one and a half standard errors. The choice of noise
generator alone changed \GateSmallSeqPosDiff{} of \GateSmallN{} sequences. Since \GateSmallN{} questions cannot
separate one- and two-point effects, we repeated the comparison on the full test split of
\GateFullN{} questions. There sequential noise answered \GateFullSeq{} correctly (\GateFullSeqPct\%), position-keyed
noise \GateFullPos{} (\GateFullSeqPosDelta{} points, standard error \GateFullSeqPosSe) and position-keyed noise with
prefix reuse \GateFullCache{} (\GateFullSeqCacheDelta{} against sequential noise, standard error
\GateFullSeqCacheSe; \GateFullPosCacheDelta{} against position-keyed noise without reuse, which served
\GateFullHitKTokens{}k prompt tokens from the cache). Both full-split variants remain within one point of sequential
noise. This result is more convincing than the subset, though it still comes from a fixed rule. As expected after a
change of random numbers, most sequences differ: only
\GateFullSeqPosSame{} of \GateFullN{} were identical between the two noise generators, and \GateFullPosCacheSame{}
between position-keyed noise with and without reuse.

\subsection{Speculation pays at low load; an adaptive budget avoids its high-load cost}

The speculative path needs fixed-depth exits and is not available for Huginn. With enough graphs that no step ran
eagerly (128, \SqFallbacks{} fallbacks), self-speculation with four draft tokens served Ouro \SqFixedOne\X,
\SqFixedTwo\X{} and \SqFixedFour\X{} the throughput of plain decode at 1, 2 and 4 requests, \SqFixedEight\X{} at 8,
\SqFixedSixteen\X{} at 16 and \SqFixedThirtyTwo\X{} at 32: once the target model saturates the GPU, the draft's extra
loops are cost. Its verification rows also multiply batch shapes; this arm captured \SqFixedGraphs{} graphs, and with
24 it ran most steps at 16 and 32 requests eagerly, which earlier made it look several times slower than here. On the
standalone host ExactLoop's speculative path was \SpfOptOne\X{} and \SpfOptEight\X{} the base engine's at 1 and 8.

The speculative path also runs a whole token, prelude, loops and coda, in one engine step (\SpkStepsZero{} steps
instead of \SpkStepsPlain{} for a 32-request burst). Without drafts (K${=}0$) that alone gave \SqZeroLowLo--\SqZeroLowHi\X{}
up to 8 requests and \SqZeroHighLo--\SqZeroHighHi\X{} at 16 and 32, with \SqZeroSame{} sequences identical at 32
(\rung{1}). A load-adaptive draft budget combines both: with at most 8 verification rows per round, each of $N$ ready
requests drafts $\min(4, \lfloor 8/N\rfloor-1)$ tokens. This gave \SqEightOne\X, \SqEightTwo\X{} and \SqEightFour\X{}
at 1, 2 and 4 requests, against \SqFixedFour\X{} for fixed $K{=}4$ at 4, and \SqEightThirtyTwo\X{} instead of
\SqFixedThirtyTwo\X{} at 32 (three runs per cell, first dropped).

\subsection{Decode is already zero-copy; copy kernels and noise syncs remain}
\label{sec:zerocopy}

We counted every host--device copy, blocking synchronization and copy kernel in a window of decode steps
(\appref{tab:census}). The engine's host side is already close to zero-copy. With asynchronous scheduling and graphs,
decode issues no host-to-device memcpy at all, because inputs are staged in host-mapped memory and read by the
kernels, and one pinned \CenDtohBytesOne-byte device-to-host copy per token, the sampled token (one batched copy
per step at 32 requests). In packed prefill, UVA staging replaces \CenPrefillHtodNouva{} host-to-device memcpys per thousand prompt
tokens (\CenPrefillHtodKBNouva~KB) with zero; without UVA staging, packed prefill ran at \UvaNouvaEight--\UvaNouvaThirtyTwo\X{} of per-row prefill
(\cref{sec:sessions}).

What remains is movement on the device and blocking reads on the host. At batch 1 the base engine launches
\CenMoveMainOne{} copy and cast kernels per token, from RMSNorm's casts and RoPE's concatenations; the exact kernel
set removes all but \CenMoveOptOne{} of them. Huginn's sampler is the other case. PyTorch~2.14's
\texttt{trunc\_normal\_} rejection-samples and checks \texttt{mask.any()} after each round, a blocking read, and the
engine calls it once per request for every token's initial state. In a Huginn decode window
ExactLoop issued \CenHugAnyOpt{} such blocking reads per step, \CenHugAnyPerTokOpt{} per output token, besides
\CenHugItemsOpt{} host-side scalar conversions. Position-keyed noise
computes the same distribution with an inverse CDF in one kernel (\cref{sec:noise}); in the same window it issued
\CenHugAnyPos{} blocking reads and \CenHugItemsPos{} scalar conversions, and device-to-host copies fell from
\CenHugDtohSeq{} to \CenHugDtohPos{} per step. In two paired runs of three repeats each, it gave Huginn
\NoisePosLoThirtyTwo--\NoisePosHiThirtyTwo\X{} the throughput of sequential noise at 32 requests,
\NoisePosLoEight--\NoisePosHiEight\X{} at 8 and \NoisePosLoOne--\NoisePosHiOne\X{} at one. On the PD
path, KV moves from the prefill pool into the decode pool's final pages with no staging copy, and with packing
only live slots move (\cref{sec:pdfixes}).

\subsection{Packed prefill and a larger in-flight budget raise PD throughput up to 3.3\X{}}
\label{sec:pd}

\begin{table}[t]
\caption{Ouro-1.4B PD on one GH200 node, base engine against the final ExactLoop tree, 64 output tokens. ExactLoop
columns give the in-flight budget (64~MB chunks for 256~MB, 256~MB chunks for 2~GB) and the UCX transport for KV:
TCP is NIXL's default error handling, IPC is CUDA IPC with error handling off. 1P1D: one prefill and one decode GPU,
2048-token prompts, three runs per arm in alternating order; 8192-token prompts and 2P2D: one or two runs. Ranges
over runs; times are medians.}
\label{tab:pdfinal}
\centering\footnotesize\setlength{\tabcolsep}{2.5pt}
\begin{tabular}{@{}lrrrr@{}}
\toprule
 & & \multicolumn{3}{c}{ExactLoop} \\
\cmidrule(l){3-5}
 & base & 256~MB, IPC & 2~GB, TCP & 2~GB, IPC \\
\midrule
\multicolumn{5}{@{}l}{\emph{1P1D, 2048-token prompts}} \\
$C{=}8$, tok/s & 107--112 & 223--228 & 228--289 & 283--289 \\
$C{=}32$, tok/s & 143--147 & 271--308 & 205--457 & 479 \\
\quad TTFT (s) & 10.8--11.1 & 3.4--3.8 & 2.0--2.1 & 2.0 \\
\quad TPOT (ms) & 30--40 & 29--57 & 21--116 & 18 \\
\quad tail (s) & 2.80--2.85 & 0.82--1.38 & 0.02 & 0.02 \\
\multicolumn{5}{@{}l}{\emph{1P1D, 8192-token prompts}} \\
$C{=}8$, tok/s & 14 & 71--80 & 89 & 79--89 \\
\quad TTFT (s) & 33.7 & 4.5--5.4 & 3.9 & 3.8--4.5 \\
\multicolumn{5}{@{}l}{\emph{2P2D, 2048-token prompts}} \\
$C{=}32$, tok/s & 236 & 446--450 & 610 & 610--617 \\
$C{=}64$, tok/s & 262 & 450--487 & 617 & 460--751 \\
\bottomrule
\end{tabular}

\end{table}

\Cref{tab:pdfinal} compares the base engine with ExactLoop on an earlier tree. With 2~GB in flight and UCX error handling off,
so that KV moves by CUDA IPC, ExactLoop served \PdfNoneLo{} tokens/s at 32 requests in all three runs against
\PdfMainLo--\PdfMainHi{} (\PdfGain\X). This arm disables UCX peer error handling, which the base arm does not, so the ratio is not a
transport-matched comparison; the error-handling arm below is the conservative one. The median TTFT fell from \PdfMainTtft{} to \PdfNoneTtft~s and the commit tail
from \PdfTailMain~s to \PdfTailNoneMs~ms. With NIXL's default error handling, which keeps peer failure detection and
runs over TCP, two runs reached \PdfPeerFastLo--\PdfPeerFastHi{} tokens/s (\PdfGainPeer\X); the third
reached \PdfPeerSlow{} with TTFT and tail unchanged and a median TPOT of \PdfPeerSlowTpotMs{} instead of
\PdfPeerFastTpotMs~ms, so it lost time in decode, not in the transfer. Averaged over all three runs, that arm is about 2.6\X{} the base; \cref{sec:lessons} traces this signature to the
CUDA graph cache. With 8192-token
prompts, \PdfLongGBPerRequest~GB of KV per request, ExactLoop reached \PdfLongNoneLo--\PdfLongNoneHi{} tokens/s against
\PdfLongMain{} and cut the median TTFT from \PdfLongMainTtft{} to \PdfLongNoneTtftLo--\PdfLongNoneTtftHi~s; an 8~GB
budget, one round per request, gave \PdfLongEightG, so at this length the budget hardly matters. With two prefill and
two decode GPUs ExactLoop served \PdfTwoNoneLo--\PdfTwoNoneHi{} tokens/s at 32 requests against \PdfTwoMain{}
(\PdfTwoOldLo--\PdfTwoOldHi{} with the old budget); at 64 requests its two runs gave \PdfTwoNoneSixtyFourLo{} and
\PdfTwoNoneSixtyFourHi, a decode-side spread like the one above. The PD and transfer-connector GPU tests passed on
both trees (\PdfTestsMain{} on the base engine, \PdfTestsPr{} on the PD branch).

Two changes produce the gain: packed prefill and the in-flight budget. In an earlier single run with the old budget, per-row prefill reached
\PdRowsTputThirtyTwo{} tokens/s at 32 requests and packed prefill \PdOptTputThirtyTwo, against \PdMainTputThirtyTwo{}
for the base engine; prefill compute fell from \PdRowsCompute{} to \PdOptCompute~s. All arms moved the same bytes,
\PdOuroKBPerToken~KB per prompt token, yet a 64~MB chunk took \PdMainXferMs~ms behind per-row prefill and
\PdOptXferMs~ms behind packed prefill: the transfer speed followed the prefill step length, which pointed at the
host rather than the link.

A single-run sweep of the budget (64~MB chunks and 256~MB in flight against 256~MB and 1~GB) separates the two
explanations. With per-row prefill, chunks four times larger took four
times longer and bandwidth stayed between \BudRowfillBigGbps{} and \BudRowfillDefaultGbps~GB/s, so the in-flight budget was
not the limit. With packed prefill, the larger budget cut the median TTFT by \BudTtftCut\%. UCX's own log showed that
intra-node remote memory access ran as active messages over TCP on the node's host network interface, and TCP
moves data only when the host polls it. That explains why chunk time grew with chunk size behind per-row
prefill, but not the whole commit tail: removing TCP left chunk times almost unchanged. With
three runs each on that tree, raising the budget from 256~MB to 2~GB raised throughput \PdfBudgetGain\X{}
(\PdfOldLo--\PdfOldHi{} to \PdfNoneLo{} tokens/s) and cut the tail from \PdfTailOld~s to \PdfTailNoneMs~ms;
\cref{sec:lessons} explains why.

\section{Lessons and Negative Results}
\label{sec:lessons}

\paragraph{Approximate fusion: too few questions.} The approximate kernels rounded once where PyTorch
rounds twice and summed in a different order, changing some elements by an ulp or two. They changed more GSM8K
sequences than the exact set and crossed our one-point rule on \GsmN{} questions, but not on the full split.
Neither sample shows that accuracy fell, or that it did not fall by a point. For looped models, a per-element ulp bound is not enough evidence of
harmlessness; the check needs enough questions to resolve the effect of interest.

\paragraph{Fused kernels cost host time in eager mode.} With fused kernels in every scope, ExactLoop ran at
\EagerPhaseOne\X{} the base engine's throughput in synchronous eager mode at batch 1 (TPOT \EagerPhaseOneTpotMain{}
to \EagerPhaseOneTpotOpt~ms, single run), and its prefill at 32 requests took \PrefillSlowOptThirtyTwo{} instead of
\PrefillSlowMainThirtyTwo~s. A Triton launch from Python costs more than the PyTorch kernels it replaces on this
host. Graph-only fusion removed the regression in prefill (\appref{tab:scope}).

\paragraph{A full graph cache turns decode eager without warning.} The engine captures one recurrent graph per batch
size, 16 by default, and runs every later size eagerly. Staggered PD imports and sessions pass through more sizes:
with 16 graphs, \GraphSixteenSlow{} of \GraphSixteenRuns{} PD runs decoded five times slower at 32 requests, and
session runs fell back to eager execution hundreds of times. ExactLoop keeps 64 graphs and evicts the least recently
used one once full; with 64 graphs no PD or session run fell back. Both changes decide which steps replay a graph,
not what they compute (\rung{0}).

\paragraph{The KV path chose TCP, and TCP was not the bottleneck.} NIXL's UCX plugin requests peer failure
handling, which shared-memory and CUDA IPC transports lack, so UCX fell back to TCP. Removing that requirement
moved the KV to CUDA IPC but barely shortened a chunk; the commit tail was set by budget rounds times the engine
step, which the larger in-flight budget removes (details in the appendix).

\section{Related Work}
\label{sec:related}

\paragraph{LLM serving.} Paged KV caching~\cite{kwon2023vllm}, iteration-level scheduling~\cite{yu2022orca}, prefix
sharing~\cite{zheng2024sglang}, and chunked prefill~\cite{agrawal2024sarathi} underpin most serving engines.
PD disaggregation~\cite{zhong2024distserve,patel2024splitwise} and KV-centric designs~\cite{qin2025mooncake} move
KV between machines; they assume one KV entry per layer, which looped models multiply by depth.

\paragraph{Variable depth and numerics.} Weight sharing across depth goes back to Universal
Transformers~\cite{dehghani2019universal}; Huginn~\cite{geiping2025huginn} and Ouro~\cite{zhu2025ouro} use it, while
CALM~\cite{schuster2022calm} and LayerSkip~\cite{elhoushi2024layerskip} also vary compute per token. Speculative
decoding~\cite{leviathan2023spec,chen2023spec} and self-speculation~\cite{zhang2024draftverify} help only while the
target model leaves the GPU idle. Batch-size-dependent kernels make outputs depend on server load
~\cite{he2025nondeterminism}; double rounding and contraction are classical~\cite{higham2002accuracy}.

\paragraph{Looped-model inference.} Continuous depth batching~\cite{schwethelm2026cdb} schedules single loop
iterations so that rows at different depths share a batch; the base engine we measure implements its runtime ideas,
and ExactLoop builds on it rather than replacing it. WaveFront~\cite{ha2026wavefront},
LoopSpec~\cite{cho2026loopspec} and LoopAhead~\cite{liu2026loopahead} draft tokens from shallow recurrent states and
verify them at full depth; FlashLoop~\cite{yang2026flashloop} skips cross-loop updates and quantizes KV residuals,
and ResidualQuant~\cite{kim2026residualquant} stores non-final loops' KV as 2-bit residuals. They report speedups
over autoregressive decoding or full-precision KV, not over a tuned engine, and change what is computed:
speculation preserves the output distribution by verification, while sparsity and quantization approximate it.
ExactLoop's \rung{0} and \rung{1} changes alter at most which rows share a batch, need no accuracy check, and are
orthogonal to these methods; we have not measured them in combination.

\paragraph{Looped models in RL.} A concurrent submission~\cite{anon2026looprl} checks learner log-probabilities
against deferred-loop rollouts on the same engine; we study serving.

\section{Conclusion}

Looped models repeat per-layer costs at every depth. Huginn wastes KV on dead depth layers; Ouro launches many
small kernels. Live-slot packing and fenced fusion remove both without changing same-batch arithmetic. Packed
prefill, position-keyed noise, graph management, and a polling-aware transfer budget supply the remaining gains. Under Poisson arrivals of real prompts, ExactLoop moves Ouro's
saturation to a higher rate but not its saturated throughput, and Huginn's ExactLoop arm had not saturated.

\bibliography{refs}
\bibliographystyle{mlsys2025}

\clearpage
\appendix
\section{Measurement Provenance and Artifact Contents}
\label{sec:artifact}

The measurements below come from several frozen revisions and jobs rather than one final tree. This appendix
records that provenance and the number of runs behind each group; \cref{tab:ledger} lists the GPU batches in
execution order. ``Retained runs'' counts runs after the warm-up policy described in the evaluation.

\begin{itemize}\setlength{\itemsep}{2pt}\setlength{\parskip}{0pt}
\item \textbf{Released defaults.} Ouro uses the paper tree. Huginn runs on an earlier tree and is checked against
the paper tree on 24 requests. Two of three runs per cell are retained. Ratios are paired within each job, but the
two models do not share one build.
\item \textbf{Plane packing.} The earlier-tree packed/unpacked comparison covers 8, 32 and 64 requests, with two
of three runs per cell retained. Same-batch identity and the changed-capacity result are reported separately.
\item \textbf{Exact kernels.} Interleaved rounds compare unfused, exact, and approximate sets, retaining four runs
per cell. Exactness is tested on the reported shapes and GPU software stack. The approximate arm is an internal
accuracy check, not a full accuracy study.
\item \textbf{Agent sessions.} Ouro and Huginn multi-turn traces retain three runs per arm after the first round,
on a tree measured before the last kernel changes.
\item \textbf{Prefix reuse identity.} The paired cold/cached check covers \CacheNoiseN{} Huginn prompts with
position-keyed noise (full hits \CacheNoiseFullSame, partial \CacheNoisePartSame) and only eight with the sequential generator.
\item \textbf{Accuracy.} The first Ouro checks use an 87-question subset; the noise, cache, packed-prefill and
repeated kernel checks use the 1319-question full split. Each arm is one deterministic greedy run; packed prefill
has a second, independent batch size for Ouro only. Packed prefill and the exact set were also checked on MATH-500 (\MathN{} questions, on the
standalone host, one run per arm; intervals reach \MathOuroPackedLo{} and \MathHugPackedLo{} points); even the GSM8K full split cannot exclude a
one-point loss for packed prefill.
\item \textbf{PD.} The 1P1D comparison uses three runs per arm; 8192-token and 2P2D checks use one or two runs.
All tests are on a single node, and the fastest IPC arm disables UCX peer error handling.
\end{itemize}

\paragraph{Artifact contents.} The repository includes the LaTeX sources, generated tables, plotted figure data,
the modified vllm-rlt snapshot, focused unit and GPU tests, and GSM8K tooling. It omits model weights, datasets,
raw profiler traces, benchmark logs, and the historical run harness. The snapshot also has later architecture
guards, so it is not the source revision for every result. Users can inspect the packed-KV mapping, prefix
coverage, loop scheduler, PD configuration, and exact kernels, and can run selected tests where the required
devices and software are available. They cannot rerun the complete performance campaign from this release.

\section{Additional Limitations}

The real-prompt replay ran on the standalone host only, with one round per arm and the last of two runs per rate;
it shares the release table's platform but not its workloads. Prompt and answer
lengths were capped at 1024 and 256 tokens, which keeps \OnShareKept{} of \OnShareRows{} ShareGPT requests.
\OnAbortRuns{} of \OnPlainRuns{} Poisson runs of Ouro's unmodified released defaults aborted with an illegal memory
access, both at 4 requests/s and neither in the same place of an exact replay. A GPU core dump of the second abort
shows a Warp MMU fault in a cuBLAS split-K GEMM (nvjet, sm\_90) right after a tensor-memory-accelerator store, that
is, a GEMM output or workspace address that was no longer mapped; ExactLoop's Triton kernels were not running. One
run each with packed prefill, fused kernels or asynchronous scheduling disabled did not abort, which at this rate
is not evidence. We suspect a tensor freed on one stream while another still used it, but the cause remains open. Our pre-registered online criterion, the highest rate whose median latency per token stays
within twice the engine's own lowest-rate value, rated ExactLoop and the base engine equal for both models, so the
predicted 1.5-fold gain in sustainable rate was not met; the criterion takes ExactLoop's lower floor as its
reference, and an absolute threshold would have been the better choice. Ouro-2.6B was measured for throughput and token identity only, not accuracy.

The released-defaults comparison, packing and kernel component tables, and final PD comparison retain two of three
runs per cell; earlier characterization and component tables often have one. At 32 requests, variation on the
standalone host overlapped several component differences, which we report as ties. Huginn's prefill share was
measured only at 32 and 64 requests and in one session run. Ouro sessions on the paper tree come from the batch-scheduled host; the release table and the Huginn sessions come
from the standalone host, where the Ouro release gains are smaller (ledger rows 48 and 52).

Huginn's GPU prefix-reuse check used \CacheNoiseN{} prompts, eight for the sequential generator. The first Ouro accuracy checks use \GsmN{} questions; the
noise, cache, packed-prefill and repeated kernel checks use the full \FullN-question split. For packed prefill the
95\% intervals still reach \FullHugPackedLo{} (Huginn) and \FullOuroPackedLo{} points (Ouro).
Bitwise kernel results apply to the tested shapes, PyTorch/CUDA versions, and GPUs. A capacity change can still
alter GEMM batch shapes and therefore the generated tokens.

PD experiments used one node because the engine starts workers locally; a cross-node path would have used TCP on
our systems. The fastest IPC result disables UCX peer error handling, leaving failures to the control channel.
Session traces contain long, repetitive prompts and may overstate cache hits elsewhere. We evaluate Ouro at depth
four and Huginn at depth 32 with fixed decoding depth. We have not measured adaptive exits, other recurrence
depths, larger models, or other GPU families.

The evaluation has no non-looped baseline of matched quality. It therefore shows how much of the looped-model overhead
the engine removes, not whether a looped model is cheaper to serve than a standard model of the same quality. Both
looped models are small, the upstream engine release is labeled experimental, and all GPU results come from one GH200
host type and one RTX~5090. Huginn packed prefill is within the one-point rule on its point estimate only (see the accuracy checks in the main text).

\section{The KV Path: TCP Was Not the Bottleneck}
\label{sec:apptcp}

NIXL's UCX plugin requests peer failure
handling by default. Shared-memory and CUDA IPC transports have no peer failure handler, so UCX falls back to TCP
for active messages and emulates remote writes with them; removing TCP from the allowed transports made the PD
engines fail at startup. With the backend created with error handling disabled, UCX's log showed CUDA IPC and
shared memory and no TCP lane, and the engines also ran with TCP removed. A 64~MB chunk still took
\UcxNotcpXferMs{} instead of \UcxPeerXferMs~ms, the commit tail stayed near one second, and behind per-row prefill
a chunk still took \UcxRowsXferMs~ms. Chunks are submitted, and their completion is observed, only between engine
steps, under a fixed in-flight budget, so the tail is set by the number of budget rounds times the step length.
That is why the larger budget helped and the faster link did not. A budget that holds one request's KV,
256~MB chunks and 2~GB in flight for \PdOuroGBPerRequest~GB per request, removes the rounds: in two runs each the commit tail fell
from \BsixNoneDefaultTailLo--\BsixNoneDefaultTailHi~s to \BsixNoneReqTailMs~ms, the median TTFT from
\BsixNoneDefaultTtftLo--\BsixNoneDefaultTtftHi{} to \BsixNoneReqTtft~s, and throughput rose to
\BsixNoneReqTputLo--\BsixNoneReqTputHi{} tokens/s. A 1~GB budget, two rounds, kept most of the
tail. Holding one request's KV is a property of this load, not a rule: with 8192-token prompts at 8 requests,
where one request's KV exceeds 2~GB, the tail was \PdfLongTailNoneMs~ms with 2~GB in flight, \PdfLongTailEightMs~ms
with 8~GB and \PdfLongTailOldLo--\PdfLongTailOldHi~s with 256~MB (the PD table of the main paper), so what counts is the KV
still in flight when prefill compute ends, against the budget. ExactLoop's default budget is now 2~GB; in-flight
bytes are not a buffer, since NIXL writes straight from the
prefill pool into the decode pool. Disabling error handling has a cost: peer failures then surface only through the
control channel.

\section{Component Tables}
\label{sec:apptables}
\setcounter{table}{0}\renewcommand{\thetable}{A\arabic{table}}

These component measurements support the evaluation section of the main paper. They come from earlier trees than the
headline tables, as each caption states.

\begin{table}[h]
\caption{Plane packing alone on Huginn-0125 with the released defaults otherwise, 512-token prompts, 64 output
tokens, GH200, same KV pool. Throughput in output tokens/s, median TTFT in seconds. Three runs per cell, first
dropped, mean of two.}
\label{tab:packing}
\centering\small\setlength{\tabcolsep}{3pt}
\resizebox{\columnwidth}{!}{
\begin{tabular}{rrrrr}
\toprule
$C$ & unpacked & packed & TTFT unpacked & TTFT packed \\
\midrule
8 & 96.5 & 97.7 & 0.60 & 0.60 \\
32 & 144.6 & 174.9 & 2.69 & 2.78 \\
64 & 133.0 & 170.9 & 10.76 & 5.48 \\
\bottomrule
\end{tabular}
}
\end{table}

\begin{table}[h]
\caption{Fused kernels on Ouro-1.4B, otherwise released defaults, 512-token prompts, 128 output tokens, standalone
GH200; speedup over all fused kernels off. Two interleaved rounds of three runs, first of each dropped, mean of four.}
\label{tab:kernels}
\centering\small\setlength{\tabcolsep}{4pt}
\begin{tabular}{rrrrr}
\toprule
$C$ & unfused (tok/s) & exact, 3-kernel norm & exact & approx. \\
\midrule
1 & 54.9 & 1.30$\times$ & 1.55$\times$ & 1.58$\times$ \\
8 & 350.2 & 1.24$\times$ & 1.59$\times$ & 1.67$\times$ \\
32 & 782.4 & 1.09$\times$ & 1.51$\times$ & 1.40$\times$ \\
128 & 1218.8 & 0.97$\times$ & 1.11$\times$ & 1.23$\times$ \\
\bottomrule
\end{tabular}

\end{table}

\begin{table}[h]
\caption{Fused-kernel scope, Ouro-1.4B, 512-token prompts, GH200, per-row prefill. ``graph'': fused kernels
only during graph capture; ``always'': every scope; ``off'': no kernel fusion; ``rerun'': a second base-engine
arm, showing run-to-run spread. Base in output tokens/s, other columns relative to it; prefill time in seconds for
base, graph and always. One kept run per cell.}
\label{tab:scope}
\centering\small\setlength{\tabcolsep}{2pt}
\resizebox{\columnwidth}{!}{
\begin{tabular}{rrrrrrr}
\toprule
$C$ & base & graph & always & off & rerun & prefill b/g/a \\
\midrule
8 & 351 & 1.38$\times$ & 1.21$\times$ & 0.97$\times$ & 1.00$\times$ & 0.44/0.33/0.57 \\
32 & 749 & 1.31$\times$ & 1.12$\times$ & 0.98$\times$ & 0.84$\times$ & 1.61/1.39/2.16 \\
64 & 927 & 1.22$\times$ & 1.09$\times$ & 1.14$\times$ & 0.96$\times$ & 4.16/3.58/4.45 \\
\bottomrule
\end{tabular}
}
\end{table}

\begin{table}[h]
\caption{Data movement per output token during decode, GH200, Ouro-1.4B, from a profiler window after prefill.
H2D/D2H: host--device memcpys (D2H~B: bytes); pageable: pageable-memory copies in the window; syncs: blocking
stream, event and device synchronizations; copy kern.: device kernels that copy or cast.}
\label{tab:census}
\centering\small\setlength{\tabcolsep}{1.5pt}
\resizebox{\columnwidth}{!}{
\begin{tabular}{rrrrrrrrr}
\toprule
decode path & engine & H2D & D2H & D2H B & pageable & syncs & copy kern. & kernels \\
\midrule
Ouro, async, $C{=}1$ & base & 0.00 & 1.00 & 8 & 0 & 2.12 & 1004.0 & 5669 \\
Ouro, async, $C{=}1$ & ExactLoop & 0.00 & 1.00 & 8 & 0 & 2.06 & 29.0 & 2852 \\
Ouro, async, $C{=}32$ & base & 0.00 & 0.06 & 8 & 0 & 0.21 & 44.2 & 249 \\
Ouro, async, $C{=}32$ & ExactLoop & 0.00 & 0.03 & 8 & 0 & 0.07 & 1.1 & 89 \\
Ouro, spec.\ $K{=}4$, $C{=}8$ & base & 0.03 & 0.03 & 13 & 0 & 0.03 & 85.5 & 480 \\
Ouro, spec.\ $K{=}4$, $C{=}8$ & ExactLoop & 0.03 & 0.03 & 13 & 0 & 0.03 & 4.4 & 249 \\
Ouro, sync, $C{=}8$ & base & 0.25 & 0.75 & 24 & 6 & 1.08 & 126.5 & 710 \\
Ouro, sync, $C{=}8$ & ExactLoop & 0.12 & 0.12 & 8 & 8 & 0.28 & 3.6 & 356 \\
Ouro, sync static, $C{=}8$ & base & 0.00 & 0.75 & 24 & 0 & 0.83 & 126.0 & 710 \\
Ouro, sync static, $C{=}8$ & ExactLoop & 0.00 & 0.12 & 8 & 0 & 0.16 & 3.5 & 356 \\
\bottomrule
\end{tabular}
}
\end{table}

\begin{table}[h]
\caption{Ouro-1.4B agent sessions, six turns, prompts up to 8000 tokens, prefix cache on, batch-scheduled GH200,
earlier tree, 64 graphs unless noted. ``rows'': per-row prefill; ``share'': per-loop prefill share. Throughput in output tokens/s,
median TTFT and TPOT in seconds. Three runs per arm, the first dropped as in the burst tables, mean of two.}
\label{tab:sessions}
\centering\small\setlength{\tabcolsep}{2pt}
\resizebox{\columnwidth}{!}{
\begin{tabular}{@{}lrrrrrr@{}}
\toprule
 & \multicolumn{3}{c}{32 sessions} & \multicolumn{3}{c}{64 sessions} \\
\cmidrule(lr){2-4}\cmidrule(l){5-7}
 & tok/s & TTFT & TPOT & tok/s & TTFT & TPOT \\
\midrule
base, 16 graphs & 98 & 0.15 & 0.30 & 72 & 1.48 & 1.08 \\
base & 113 & 0.15 & 0.20 & 78 & 1.95 & 0.99 \\
ExactLoop, no share, rows & 135 & 0.34 & 0.20 & 91 & 3.44 & 0.43 \\
\quad + packed prefill & 187 & 0.13 & 0.12 & 218 & 0.94 & 0.14 \\
released (+ share) & 179 & 0.11 & 0.12 & 192 & 0.25 & 0.32 \\
\bottomrule
\end{tabular}
}
\end{table}

\begin{table*}[t]
\caption{Ledger of GPU batches, in the order run. ``5090'': rented RTX~5090; ``GH200-B'': batch-scheduled GH200 (one GPU,
two for PD); ``GH200-S'': standalone GH200 host. Refuted expectations and batches without data are included.}
\label{tab:ledger}
\centering\scriptsize
\renewcommand{\arraystretch}{0.92}
\setlength{\tabcolsep}{2.5pt}
\begin{tabular}{@{}rlp{0.3\textwidth}p{0.585\textwidth}@{}}
\toprule
\# & system & question & outcome \\
\midrule
1 & 5090 & Where does time go in the base engine? & Huginn's pool holds \HugFiveKvTokens{} tokens and throughput stays near \HugFiveTputCEight{} tokens/s from 8 requests on; speculative decoding wins only at low load; hit rates follow pool size. \\
2 & 5090 & Do fewer submissions speed up decode? & Submissions per token \FiveLoopSubsMain{} to \FiveLoopSubsOpt; throughput \FiveLoopGainOne\X{} at batch 1 and \FiveLoopGainThirtyTwo\X{} at 32 requests; \FiveLoopSame{} requests identical. \\
3 & 5090 & Where does host time go? & Per-row KV bookkeeping is visible at 32 requests; a cheaper shared record broke the per-layer safety contract and was withdrawn. \\
4 & 5090 & Can PRELUDE be fused into decode? & Regression: batches split into small cohorts. A policy fix removed it, the remaining gain was small, and the feature stays opt-in. \\
5 & 5090 & Do cache fixes and packed prefill help sessions? & Packed prefill was the largest session lever; the hit rate tracked pool size. \\
6 & 5090 & How much do fused norm, RoPE and SiLU kernels give? & Large batch-1 gain; greedy outputs changed. \\
7 & 5090 & Which fused kernel carries the gain? & RMSNorm. \\
8 & 5090 & Are approximate kernels harmless on GSM8K? & Both missed the one-point rule: \GsmNormDelta{} and \GsmAllDelta{} points on \GsmN{} questions. That subset cannot show whether accuracy drops. Loop fusion: \GsmNoneSeqDiff{} of \GsmN{} sequences differ. \\
9 & 5090 & Is loop fusion identical on real text at low load? & \FiveRealLoop{} sequences identical; two base runs \FiveRealSelf. \\
10 & GH200-B & Is plane packing bitwise, and how much capacity does it add? & \PackSame{} identical with the same batch; \PfSame{} when the larger capacity changes batch composition. \PackCapacityRatio\X{} tokens, \PackGainThirtyTwo\X{} and \PackGainSixtyFour\X{} at 32 and 64 requests. Session arms exceeded Huginn's context: no data. \\
11 & GH200-B & Is the fused RMSNorm bitwise? & No: its residual epilogue rounded once, and \DiagDouble\% of elements differed by one ulp. Fixed with PyTorch's addition. \\
12 & GH200-B & Is the fixed norm bitwise; does packed prefill help long prompts? & All requests identical; packed prefill \PackedPtwokEight\X{} and \PackedPtwokThirtyTwo\X{} with 2048-token prompts. \\
13 & GH200-B & Does packed prefill work for Huginn? & Yes, up to \HugPackedPtwokGainThirtyTwo\X; tokens change, so it needed a gate. \\
14 & GH200-B & Can $\mathrm{rsqrt}$ move into the tail kernel bitwise? & Yes; outputs identical with fewer launches. \\
15 & GH200-B & Do fenced RoPE and SiLU stay bitwise? & \ExactSame{} identical; \ExactGainOneTwoEight--\ExactGainOne\X. \\
16 & GH200-B & Four-GPU end-to-end batch & Cancelled before it started, at the user's request; its arms moved to GH200-S. \\
17 & GH200-S & Do the exact kernels and Huginn packed prefill pass accuracy gates? & Ouro: \GateOuroSeqDiff{} of \GsmN{} sequences differ. Huginn: \HugGateRows{} vs \HugGatePacked{} correct of \HugGateN. SiLU: \SiluMismatch{} mismatching elements. \\
18 & GH200-S & How much faster is ExactLoop end to end? & Huginn \HugOptMin--\HugOptMax\X. Ouro eager mode \EagerPhaseOne\X: fused kernels outside graphs cost host time. Session arms timed out: no data. \\
19 & GH200-S & Does graph-only fusion fix that? What happens in sessions? & Prefill regression gone. Loop fusion delayed admission (median TTFT \SessMainTtftSixtyFour{} to \SessOptTtftSixtyFour~s). Huginn prefix cache \HugSessIsoGain\X{} (single run). \\
20 & GH200-S & Is packed prefill slower for short prompts? & No; an earlier reading came from a slow first repeat and was retracted. Without UVA staging, \UvaNouvaEight\X{} and \UvaNouvaThirtyTwo\X. \\
21 & GH200-S & Does the per-loop prefill share undo the TTFT cost? & Bursts: better on every metric. Sessions: TTFT down, time per token up. \\
22 & GH200-S & Released defaults, repeats; GPU cache identity; census; ncu & \VtwoOuroShortMin--\VtwoOuroShortMax\X{} (Ouro), \VtwoHugMin--\VtwoHugMax\X{} (Huginn) in every repeat. Cache hits match cold runs in \CacheNoiseFullSameEight{} cases. Decode is zero-copy on the host. No ncu: counters restricted. \\
23 & GH200-B & First PD runs & No data: a benchmark bug, an upstream coordinator that could not start Huginn, and an overlap counter that missed coalesced transfers. \\
24 & GH200-B & How much does ExactLoop help PD? & \PdOptGainThirtyTwo\X{} at 32 requests; commit tail as long as prefill compute; Huginn packing slower in one run. \\
25 & GH200-B & Is the transfer limited by the budget or by the link? & Not the budget behind per-row prefill; with packed prefill a larger budget cut TTFT by \BudTtftCut\%. UCX used TCP. \\
26 & GH200-B & Does KV move faster without TCP? & The engines failed at startup: peer error handling needs TCP. \\
27 & GH200-B & Same, with error handling off & TCP gone, but a chunk took \UcxNotcpXferMs{} instead of \UcxPeerXferMs~ms: our TCP explanation was wrong. \\
28 & GH200-B & Larger budgets under CUDA IPC; Huginn PD packing repeated & A 2~GB budget, one request's KV, cut the commit tail to \BsixNoneReqTailMs~ms and gave \BsixNoneReqTputLo--\BsixNoneReqTputHi{} tokens/s in both runs; now the default. Huginn packing in PD: no slowdown in two repeats. \\
29 & GH200-S & Does position-keyed noise cost speed? & No: \NoisePosLoThirtyTwo--\NoisePosHiThirtyTwo\X{} of sequential noise at 32 requests, ties at 1 and 8; it removes Huginn's blocking noise reads. \\
30 & GH200-S & Component effects and a Huginn gate, earlier tree, with repeats & Exact kernels \KfVfourExactMin--\KfVfourExactMax\X{}, \KfVfourExactSame{} identical; packing \PfGainThirtyTwo\X{} and \PfGainSixtyFour\X{} at 32 and 64 requests; Huginn GSM8K \GateSmallSeq, \GateSmallPos{} and \GateSmallCache{} of \GateSmallN{} (sequential, position, position with reuse). Base-engine arm and branch tests: no data (bad flag; no GPU flag). \\
31 & GH200-B & Final PD with repeats; PD tests; ncu & Base engine \PdfMainLo--\PdfMainHi{} tokens/s, three runs. ExactLoop ran at the old budget (the benchmark's default overrode the engine's); tests ran without the GPU. ncu: \NcuMinUs--\NcuMaxUs~$\mu$s per exact kernel. \\
32 & GH200-S & Same, fixed: base-engine Huginn arm; branch GPU tests & Final tree \HsplitFullOne--\HsplitFullThirtyTwo\X{} the base engine on Huginn, \HsplitRestOne--\HsplitRestThirtyTwo\X{} without fused kernels. No branch adds a GPU test failure: each failure is a missing FlashAttention package or also fails on the base engine. \\
33 & GH200-B & Same, fixed: 2~GB budget, repeats, PD GPU tests & CUDA IPC \PdfGain\X{} at 32 requests (\PdfNoneLo{} vs \PdfMainLo--\PdfMainHi{} tokens/s), TCP \PdfGainPeer\X{} in two of three runs; 8192-token prompts \PdfLongNoneLo--\PdfLongNoneHi{} vs \PdfLongMain; PD tests pass. \\
34 & GH200-S & Huginn gate on the full test split & No data at first (the protocol step needs a git checkout). Rerun: \GateFullSeq, \GateFullPos, \GateFullCache{} of \GateFullN{} (sequential, position, with reuse): pass. \\
35 & GH200-S & Same, with FlashAttention 3 and 4 & FlashAttention tests pass; two failures remain, also on the base engine. \\
36 & GH200-B & Adaptive drafts; the slow TCP run & K${=}0$ \SpkZeroLo--\SpkZeroHi\X; the slow decode recurred once in five runs. \\
37 & GH200-B & Speculation crossover; graph counters & Budget 8 kept \SpkEightOne\X{} at one request; the slow run had \GraphSlowFallbacks{} eager graph fallbacks. \\
38 & GH200-B & Does graph eviction remove the cliff? & PD at budget 4: \GraphFourOldLo--\GraphFourOldHi{} to \GraphFourNew{} tokens/s; sessions gained, GPU tests pass. \\
39 & GH200-B & How many graphs by default? & 64: sessions never fell back; six PD runs neither. \\
40 & GH200-B & Sessions again: three runs, graph counters & Base 16 to 64 graphs \SsGraphGainThirtyTwo\X{} and \SsGraphGainSixtyFour\X; released \SsRelGainThirtyTwo\X{} and \SsRelGainSixtyFour\X{} the base engine; Huginn \HugSessPdThirteenGain\X; earlier single runs superseded. \\
41 & GH200-B & Prefill share in sessions, three runs & Share cut TTFT \SsPackedTtftSixtyFour{} to \SsRelTtftSixtyFour~s at \SsShareGainSixtyFour\X{} throughput; no share with per-row prefill \SsRowsTtftSixtyFour~s. \\
42 & GH200-B & Release table and speculation with enough graphs & No fallbacks: Ouro \RelOldOuroShortMin--\RelOldOuroShortMax\X{} and \RelOldOuroLongMin--\RelOldOuroLongMax\X, Huginn \RelOldHugMin--\RelOldHugMax\X; fixed $K{=}4$ \SqFixedThirtyTwo\X{} at 32, not \SpkFixedThirtyTwo\X. \\
43 & GH200-B & FlashAttention 3 and 4 instead of Triton & FA3 \FaThreeLo--\FaThreeHi\X, FA4 \FaFourLo--\FaFourHi\X{} of Triton: no gain. \\
44 & GH200-B & Prefill split; exact eager kernels & Elementwise: half of Ouro prefill; fused $-$\RowsOuroPrefillCut\%, identical. \\
45 & GH200-B & Huginn act/RoPE; one-kernel norm & \HugFuseLo--\HugFuseHi\X, \OneHugLo--\OneOuroShortHi\X, bitwise; first session run slow. \\
46 & GH200-B & Release table, one-kernel norm tree & Ouro \RelPrevOuroShortMin--\RelPrevOuroShortMax\X{} and \RelPrevOuroLongMin--\RelPrevOuroLongMax\X, Huginn \RelPrevHugMin--\RelPrevHugMax\X. \\
47 & GH200-S & Kernel table, one-kernel norm, interleaved & One-kernel norm: exact \KfExactMin--\KfExactMax\X, approx.\ \KfApproxMin--\KfApproxMax\X; \KfExactSame{} identical. \\
48 & GH200-B & Tree before fused projections; sessions interleaved; order probe & Ouro \RelSixOuroShortMin--\RelSixOuroShortMax\X, sessions \SsSixGainThirtyTwo--\SsSixGainSixtyFour\X; probe \ProbeCases/\ProbeCases{} bitwise. \\
49 & GH200-B & One GEMM for q/k/v and gate/up; paper tree & Bitwise; \ProjAllLo--\ProjAllHi\X{} (rule set before the run), adopted: Ouro \RelBOuroShortMin--\RelBOuroLongMax\X, sessions \SsFinalGainThirtyTwo--\SsFinalGainSixtyFour\X. \\
50 & GH200-S & Do the accuracy decisions hold on the full split? & Exact = unfused on all \FullN; approx.\ \FullOuroApproxDelta{} and \FullOuroApproxNormDelta; packed: Ouro \FullOuroPackedDelta{} (batch 16: \FullOuroPackedDeltaBSixteen), Huginn \FullHugPackedDelta; batch 64 replays batch 32. \\
51 & GH200-S & Is Huginn's packed result a sampling accident? How often do GPU cache hits match? & Batch 16 replays batch 32 (\FullHugSixteenSame/\FullN{} identical): one observation. Position-keyed noise, \CacheNoiseN{} prompts: full hits \CacheNoiseFullSame, partial \CacheNoisePartSame. \\
52 & GH200-S & Do the release and Huginn-session gains hold with one build? & Same token identity; Ouro \RelOuroShortMin--\RelOuroShortMax\X{} (512) and \RelOuroLongMin--\RelOuroLongMax\X{} (2048), Huginn \RelHugMin--\RelHugMax\X; sessions \HugSessGain\X. Smaller Ouro gains than GH200-B; the release and Huginn-session tables use this run \\
53 & GH200-S & Real prompts, Poisson arrivals, a third model? & ShareGPT offline Ouro \OnOuroOffShareGPTGain\X, Huginn \OnHugOffShareGPTGain\X; online latency per token up to \OnOuroNlGainMax\X{} and \OnHugNlGainMin--\OnHugNlGainMax\X{} lower; Ouro saturates later at the same ceiling; pre-registered sustainable-rate criterion not met. Ouro-2.6B \OuroBigShortMin--\OuroBigLongMax\X. \OnAbortRuns{} of \OnPlainRuns{} Ouro runs aborted (cuBLAS split-K GEMM, MMU fault). \\
54 & GH200-S & Does packed prefill hold on a second benchmark? & MATH-500, \MathN{} questions: Ouro \MathOuroPackedDelta{} (s.e.\ \MathOuroPackedSe), Huginn \MathHugPackedDelta{} (\MathHugPackedSe); exact = unfused on \MathOuroUnfusedSame{} of \MathN. \\
\bottomrule
\end{tabular}
\end{table*}

\end{document}